\documentclass[prl,aps,showpacs,twocolumn,preprintnumbers,amsmath,amssymb,superscriptaddress,longbibliography,nofootinbib]{revtex4-2}
\usepackage[english]{babel}
\usepackage{amsmath,amssymb,bbm,mathrsfs,mathtools,multirow,diagbox,xcolor,graphicx,tabularx,comment,amsfonts,dsfont,lettrine,units,comment}
\pdfoutput=1

\usepackage{graphicx}
\usepackage[colorlinks=True,linkcolor=red,citecolor=blue,urlcolor=blue]{hyperref}
\usepackage{bookmark}
\usepackage{xcolor}
\usepackage{chngcntr}
\usepackage{braket}
\usepackage{nicefrac}
\usepackage{bm}
\usepackage{bbm}
\usepackage{braket}
\usepackage{lineno}
\usepackage{array}
\newcolumntype{C}{>{$}c<{$}}
\usepackage{newtxtext} 
\usepackage[normalem]{ulem}
\usepackage{wasysym}
\usepackage{cancel}

\usepackage{amsmath}
\newcommand{\angstrom}{\textup{\AA}}

\usepackage{graphicx}
\usepackage{bm}

\usepackage{hyperref}
\hypersetup{
    colorlinks=true,
    linkcolor=blue,
    filecolor=blue,      
    urlcolor=blue,
    citecolor=blue,
    pdftitle={spillage},
    pdfauthor={Daniel Muñoz-Segovia},
    pdfkeywords={spillage},
}
 
\makeatletter
\renewcommand*\env@matrix[1][*\c@MaxMatrixCols c]{%
  \hskip -\arraycolsep
  \let\@ifnextchar\new@ifnextchar
  \array{#1}}
\makeatother

\begin{document}

\title{Structural spillage:\\ 
an
efficient method to identify non-crystalline topological materials}

\author{Daniel Mu\~noz-Segovia\textsuperscript{*}}%
\email{daniel.munozsegovia@dipc.org}
\affiliation{Donostia International Physics Center, 20018 Donostia-San Sebastian, Spain}
\affiliation{Univ.~Grenoble Alpes, CNRS, Grenoble INP, Institut Néel, 38000 Grenoble, France}
\author{Paul Corbae\textsuperscript{*}}
\affiliation{Department of Materials Science, University of California, Berkeley, California 94720, USA}
\affiliation{Materials Sciences Division, Lawrence Berkeley National Laboratory, Berkeley, California, 94720, USA}
\author{D\'aniel Varjas}
\affiliation{Department of Physics, Stockholm University, AlbaNova University Center, 106 91 Stockholm, Sweden}
\affiliation{Max Planck Institute for the Physics of Complex Systems, N{\"o}thnitzer Strasse 38, 01187 Dresden, Germany}
\author{Frances Hellman}
\affiliation{Department of Physics, University of California, Berkeley, California 94720, USA}
\affiliation{Materials Sciences Division, Lawrence Berkeley National Laboratory, Berkeley, California, 94720, USA}
\author{Sinéad M. Griffin}
\email{sgriffin@lbl.gov}
\affiliation{Materials Sciences Division, Lawrence Berkeley National Laboratory, Berkeley, California, 94720, USA}
\affiliation{Molecular Foundry Division, Lawrence Berkeley National Laboratory, Berkeley, California, 94720, USA}
\author{Adolfo G. Grushin}%
\email{adolfo.grushin@neel.cnrs.fr}
\affiliation{Univ.~Grenoble Alpes, CNRS, Grenoble INP, Institut Néel, 38000 Grenoble, France}

\date{\today}

\begin{abstract}
While topological materials are not restricted to crystals, there is no efficient method to diagnose topology in non-crystalline solids such as amorphous materials. Here we introduce the structural spillage, a new indicator that predicts the unknown topological phase of a non-crystalline solid, which is compatible with first-principles calculations. We illustrate its potential with tight-binding and first-principles calculations of amorphous bismuth, predicting a bilayer to be a new topologically nontrivial material. Our work opens up the efficient prediction of non-crystalline solids via first-principles and high-throughput searches. 
\end{abstract}

\maketitle

\paragraph*{Introduction-.} 

Predicting which solids host non-trivial electronic topological phases is a central problem in condensed matter physics. For crystalline solids, first principles methods take advantage of crystal symmetries to identify topological materials~\cite{Kruthoff17,Po:2017ci,Song2018,Frey_et_al:2020,Wieder22}. However, symmetry-based methods cannot be applied to diagnose non-trivial topology in materials that lack translational invariance such as amorphous, polycrystalline, and quasicrystalline materials. In fact, given the far greater ubiquity of non-crystalline materials in condensed matter, solving this challenge would open up several new material classes far more numerous than crystals, with both fundamental interest for novel phenomena unique to non-crystalline matter~\cite{Prodan2013,Agarwala:2017jv,Mitchell2018,Poyhonen2017,Bourne:2018jr,corbae_evidence_2021,yang_topological_2019,costa_toward_2019,Marsal2020,Spring2021,Marsal2022,Sahlberg20,Grushin2020,agarwala_higher-order_2020,mukati_topological_2020,Spring2021,Marsal2022,wang_structural-disorder-induced_2021,wang_structural_2022,manna_noncrystalline_2022,corbae_structural_2021,kraus_topological_2012,mei_simulating_2012,kraus_four-dimensional_2013,madsen_topological_2013,verbin_observation_2013,deng_topological_2014,tran_topological_2015,bandres_topological_2016,fulga_aperiodic_2016,lau_topological_2016,Varjas2019,Zilberberg:21}, and for their possible greater ease of integration into devices~\cite{Zallen,leGallo2020overview}.

Prior work on topology in non-crystalline materials used convenient amorphous tight-binding models with average and local symmetries~\cite{Marsal2020,Spring2021,corbae_evidence_2021,Marsal2022,uria-alvarez_deep_2022}, however these  do not include the full chemical and structural specificity found in real matter. Similarly, real-space invariants~\cite{kitaev_anyons_2006,prodan_non-commutative_2010,bianco_mapping_2011,hannukainen_local_2022}, including Wannier-based tight-binding formalism, require the system be treated on a case-by-case basis and can be computationally costly. 

\begin{figure}
    \centering
    \includegraphics[width = \columnwidth]{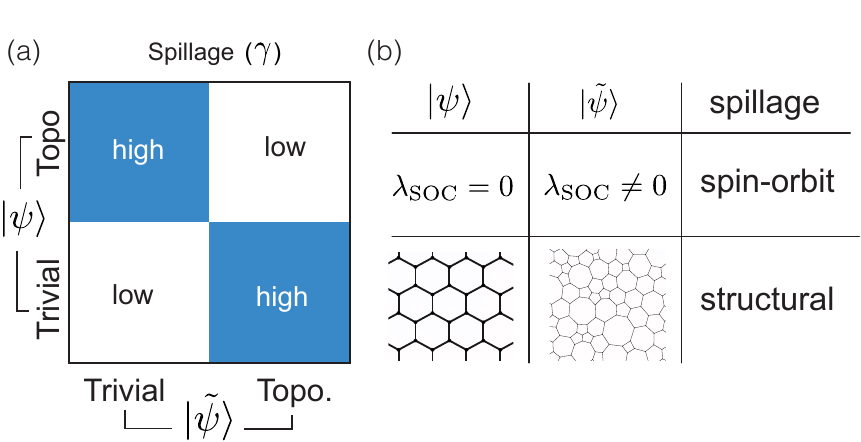}
    \caption{(a) The spillage $\gamma$ is high or low depending on whether a test wavefunction $|\psi\rangle$ is in the same or different topological state compared to a known reference wavefunction $|\tilde{\psi}\rangle$. 
    (b) The spin-orbit spillage~\cite{liu_spin-orbit_2014} compares wavefunctions with and without SOC. The structural spillage takes advantage of the knowledge of 
    the topological state of a crystalline solid to find the topological state of an amorphous solid.
    }
    \label{fig:fig1}
\end{figure}

To overcome this methodological problem, we introduce the `structural spillage', which is inherently compatible with first-principles approaches.  Since the characterization of topology in general relies on the comparison with a known reference~\cite{Griffin/Spaldin:2017}, we propose that in our case the appropriate comparison is between the wavefunctions of the non-crystalline target system and a crystalline reference state. A similar approach was proposed to identify topological band inversions in crystals by Liu and Vanderbilt~\cite{liu_spin-orbit_2014} who compared the wavefunction overlap in crystals with and without spin-orbit coupling (the `spin-orbit' spillage). Inspired by this idea, we  define the structural spillage as a measure of the overlap between wavefunctions with different structural configurations. By comparing this structural spillage for crystals, whose topological characterization can be efficiently calculated using standard symmetry-based methods~\cite{Kruthoff17,Po:2017ci,Song2018,Frey_et_al:2020,Wieder22}, with those of non-crystalline solids, the topological characterization of the latter can be determined (Fig.~\ref{fig:fig1}).

We first define the general formulation of structural spillage and how it can be used to diagnose topology in non-crystalline systems once a known reference phase is identified. We next exemplify its potential by diagnosing topological phase transitions in amorphous bismuth, a previously identified non-trivial amorphous system, using both a tight-binding model and density functional theory (DFT). Our results indicate that the structural spillage can accurately identify amorphous bismuthene as topologically non-trivial~\cite{costa_toward_2019,focassio_structural_2021}, and predicts that amorphous bilayer bismuth is a novel topological material. By definition, the structural spillage is applicable to generic non-crystalline materials. It is suitable to establish a high-throughput catalogue of potential non-crystalline topological materials, using currently available DFT codes based on plane waves in our current formalism.

\paragraph*{Structural spillage-.}
The total spillage $\gamma$ measures the mismatch between two projectors $P$ and $\tilde{P}$ into occupied states~\cite{liu_spin-orbit_2014}
\begin{equation}
    \label{eq:totalspillage}
	\gamma = \frac{1}{2} \text{Tr} \left[ \left(P - \tilde{P} \right)^2 \right] = \text{Tr} \left[ P (1-\tilde{P}) \right],
\end{equation}
where the trace acts on the entire Hilbert space, and the last equality holds under the assumption that both systems have the same total number of occupied states $N_{\mathrm{occ}} = \mathrm{Tr}[P] = \mathrm{Tr}[\tilde{P}]$. By definition, $\gamma \geq 0$ and can be viewed as the variance between two distributions with the same average. When $P=\tilde{P}$ the spillage vanishes. However, when the overlap between the two projectors is zero, it equals the total number of occupied states $N_{\mathrm{occ}}$. Therefore, $\gamma$ acts as an indicator of band inversions caused by the parameters that differ in $P$ and $\tilde{P}$~\cite{liu_spin-orbit_2014}.

To predict topological band inversions in crystals, Liu and Vanderbilt \cite{liu_spin-orbit_2014} chose $P$ and $\tilde{P}$ to be projectors onto the subspace of occupied states of crystalline insulators with and without spin-orbit coupling (SOC), respectively. Lattice periodicity allows these to be written in Bloch momentum $\boldsymbol{k}$ as $P(\boldsymbol{k}) = \sum_{n \in \text{occ}} \left| \psi_{n \boldsymbol{k}} \rangle \langle \psi_{n \boldsymbol{k}} \right|$, which defines a $\boldsymbol{k}$-resolved spin-orbit Bloch spillage, $\gamma_{\mathrm{B}}(\boldsymbol{k}) = n_{\text{occ}} - \mathrm{Tr} [ P(\boldsymbol{k}) \tilde{P}(\boldsymbol{k}) ]$, where $n_{\text{occ}} = N_{\text{occ}} / N_{\text{cells}}$ is the number of occupied bands. The total spillage is recovered by summing over all momenta in the Brillouin zone (BZ), $\gamma = \sum_{\boldsymbol{k}} \gamma_{\mathrm{B}}(\boldsymbol{k})$. The spin-orbit Bloch spillage $\gamma_{\mathrm{B}}(\boldsymbol{k})$ thus quantifies the band inversion caused by SOC at each $\boldsymbol{k}$; it is large at points in the BZ where the band inversion is sizable. Ref.~\cite{liu_spin-orbit_2014} showed that at certain points in the BZ the spin-orbit Bloch spillage has to be larger than some given value if the SOC induces a topologically non-trivial phase from Wannier obstruction arguments. For instance, this lower bound equals two for a time-reversal symmetric topological insulator.

From the above properties, $\gamma_{\mathrm{B}}(\boldsymbol{k})$ can be used to signal topological band inversions in crystals, and is straight-forward to calculate using DFT~\cite{liu_spin-orbit_2014}. Indeed, it has recently been applied to high-throughput searches for topological crystals \cite{choudhary_high-throughput_2019,choudhary_high-throughput_2021}. We note, however, that a large spillage is a necessary but not sufficient condition for non-trivial topology: in certain cases, e.g., when many bands close to the Fermi level are slightly mixed by SOC, the spillage may be fooled by trivial insulators~\cite{liu_spin-orbit_2014}. Consequently, more recent searches for topological crystals favor symmetry-based methods. In most practical cases, the spillage is expected to be an accurate indicator of topology in crystals~\cite{liu_spin-orbit_2014}. 

In this work, we propose a spillage that compares an amorphous system with a crystalline counterpart. In doing so, we take advantage of the well-developed methods of symmetry indicators for the topological characterization of crystals~\cite{Po:2017ci}. To this end, we now reformulate our spillage in a plane-wave basis for incorporation into standard plane-wave DFT codes. Moreover, it is also well defined for both crystalline and non-crystalline systems. We write the total spillage $\gamma$ in the plane wave basis $|\boldsymbol{p} \alpha \rangle$, where $\boldsymbol{p}$ is the plane-wave momentum (not necessarily restricted to the first BZ) and $\alpha$ denotes spin. To calculate the spillage, we need the projector onto occupied states of the amorphous and reference systems, $P = \sum_{N\in\mathrm{occ}} |\psi_N\rangle\langle\psi_N|$, where $|\psi_N\rangle$ are the eigenstates. By projecting these onto plane waves, we then have access to the projector matrix elements $P^{\alpha\beta}_{\boldsymbol{p},\boldsymbol{p}'} = \left\langle \boldsymbol{p} \alpha \right| P \left| \boldsymbol{p}' \beta \right\rangle$, which are well-defined for crystalline and non-crystalline systems. Any plane-wave momentum $\boldsymbol{p}$ can be uniquely decomposed as $\boldsymbol{p} = \boldsymbol{k}+\boldsymbol{G}$, the sum of a crystal momentum $\boldsymbol{k}$ in the first BZ plus a reciprocal lattice vector $\boldsymbol{G}$, both of the reference crystal. Then, by substituting the plane-wave expansion into Eq.~\eqref{eq:totalspillage}, we can define the quasi-Bloch spillage as 
\begin{widetext}
\begin{subequations} 
\begin{align}
	\gamma_{\mathrm{qB}}(\boldsymbol{k}) & = \frac{1}{2} \sum_{\boldsymbol{k}'} \sum_{\boldsymbol{G}\boldsymbol{G}'} \sum_{\alpha\beta} \left[ P^{\alpha\beta}_{\boldsymbol{k}+\boldsymbol{G},\boldsymbol{k}'+\boldsymbol{G}'} P^{\beta\alpha}_{\boldsymbol{k}'+\boldsymbol{G}',\boldsymbol{k}+\boldsymbol{G}} - P^{\alpha\beta}_{\boldsymbol{k}+\boldsymbol{G},\boldsymbol{k}'+\boldsymbol{G}'} \tilde{P}^{\beta\alpha}_{\boldsymbol{k}'+\boldsymbol{G}',\boldsymbol{k}+\boldsymbol{G}} \right] + \left[P \leftrightarrow \tilde{P} \right] = \label{eq:qB-spillage-general} \\
	& = \frac{1}{2} \left\{ \left[ \sum_{\boldsymbol{G} \alpha} P^{\alpha\alpha}_{\boldsymbol{k}+\boldsymbol{G},\boldsymbol{k}+\boldsymbol{G}} \right] + \tilde{n}_{\text{occ}}(\boldsymbol{k}) - \sum_{\boldsymbol{G} \alpha} \sum_{\boldsymbol{G}'\beta} \left[ P^{\alpha\beta}_{\boldsymbol{k}+\boldsymbol{G},\boldsymbol{k}+\boldsymbol{G}'} \tilde{P}^{\beta\alpha}_{\boldsymbol{k}+\boldsymbol{G}',\boldsymbol{k}+\boldsymbol{G}} + \tilde{P}^{\alpha\beta}_{\boldsymbol{k}+\boldsymbol{G},\boldsymbol{k}+\boldsymbol{G}'} P^{\beta\alpha}_{\boldsymbol{k}+\boldsymbol{G}',\boldsymbol{k}+\boldsymbol{G}} \right] \right\} \label{eq:qB-spillage-crystal}
\end{align}
\label{eq:qB-spillage}
\end{subequations}
\end{widetext}
In Eq.~\eqref{eq:qB-spillage-crystal} we have used the fact that the reference projector $\tilde{P}$ corresponds to a crystal, which allows us to set $\boldsymbol{k}' = \boldsymbol{k}$ in terms involving at least one $\tilde{P}$, since there is no scattering between different crystal momenta due to the discrete translational symmetry. Note that $\gamma_{\mathrm{qB}}(\boldsymbol{k})$ fulfills the same sum rule as the Bloch spillage, $\gamma = \sum_{\boldsymbol{k}} \gamma_{\mathrm{qB}}(\boldsymbol{k})$. Therefore, applied to two insulating crystals, $\gamma_{\mathrm{qB}}(\boldsymbol{k})$ recovers the Bloch spillage. Moreover, it can also be applied to semimetallic systems with the advantage of it being bounded by zero, in contrast to recent extensions to semimetallic materials \cite{choudhary_high-throughput_2019,choudhary_high-throughput_2021}. 

Our key result is that the structural quasi-Bloch spillage, defined by Eq.~\eqref{eq:qB-spillage}, can be used as an efficient topological indicator in non-crystalline systems. Crucially, it can be efficiently computed with plane-wave-based DFT methods, since the projector matrix elements are an output of the calculation. Consequently, this method is suitable for high-throughput identification of non-crystalline topological materials.

\begin{figure}
    \centering
    \includegraphics[width = \columnwidth]{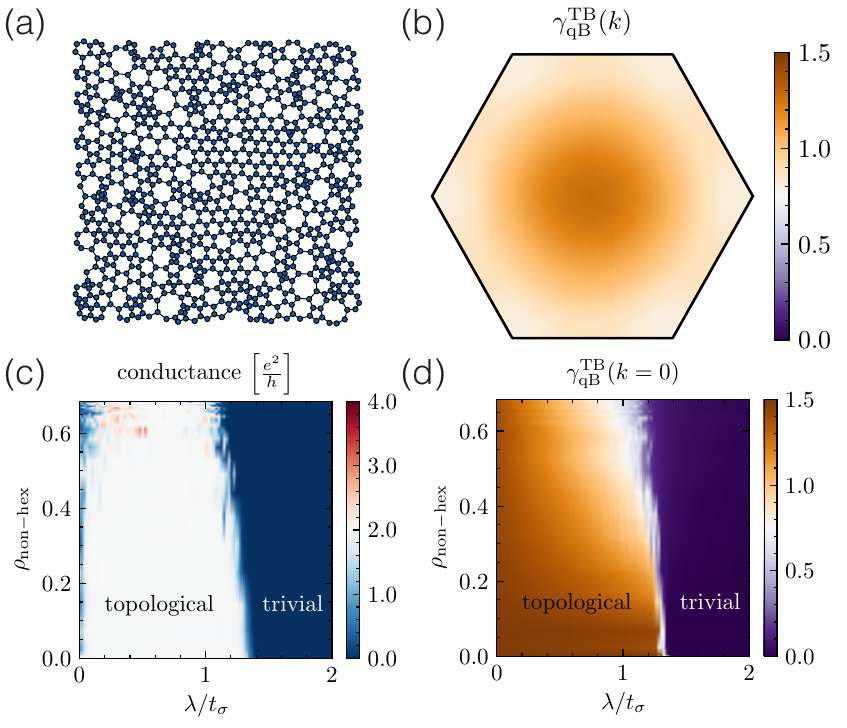}
    \caption{Structural spillage in the tight-binding approximation. 
    (a) Example of a real-space structure with a density of non-hexagonal plaquettes $\rho_{\text{non-hex}} \simeq 0.53$. 
    (b) Structural quasi-Bloch spillage $\gamma_{\mathrm{qB}}^{\mathrm{TB}}(\boldsymbol{k})$ in the BZ comparing topological amorphous bismuthene with {$\rho_{\text{non-hex}} \simeq 0.53$} and $\lambda = 0.22 t_{\sigma}$ with a trivial crystal with $\lambda/t_{\sigma}=\infty$. 
    (c), (d) Phase diagrams as a function of SOC $\lambda$ and the density of non-hexagonal plaquettes $\rho_{\text{non-hex}}$. 
    (c) Conductance in the ``armchair'' ribbon configuration (see SM~\cite{SuppMat} \ref{app:TB_details}).
    (d) Structural quasi-Bloch spillage $\gamma_{\mathrm{qB}}^{\mathrm{TB}}(\boldsymbol{k}=0)$ comparing the amorphous system to a trivial crystal with $\lambda/t_{\sigma}=\infty$.}
    \label{fig:fig_panel}
\end{figure}

\paragraph*{Structural spillage in the tight-binding approximation-.}
Defining a structural spillage that is useful in the tight-binding approximation requires us to develop further Eq.~\eqref{eq:qB-spillage}. The reason is that two issues emerge as we define plane wave states projected into the tight-binding Hilbert space of $N_{\text{sites}}$ as $\big| \boldsymbol{p} \alpha \rangle = \frac{1}{\sqrt{N_{\text{sites}}}} \sum_{\boldsymbol{r}} e^{i \boldsymbol{p} \cdot \boldsymbol{r}} \big| \boldsymbol{r} \alpha \rangle$, where $\boldsymbol{r}$ labels the position of each site and $\alpha$ labels internal quantum numbers, such as spin or the orbital type. First, because the tight-binding model's Hilbert space does not span the entire real space but only positions defined by the charge centers, our plane waves are non-orthogonal. Therefore, their overlap depends on the atomic positions, and therefore on the amount of structural disorder. Since we expect continuous translational symmetry to be recovered after averaging over different disorder realizations, we may solve this issue by neglecting the scattering between different momenta in Eq.~\eqref{eq:qB-spillage}, \textit{i.e.} assuming that $P^{\alpha\beta}_{\boldsymbol{p},\boldsymbol{p}'} \propto \delta_{\boldsymbol{p},\boldsymbol{p}'}$. This assumption has been successfully used to determine the topology of non-crystalline systems using the effective Hamiltonian approach \cite{Varjas2019,Marsal2020,Spring2021,Marsal2022}.

A second issue of the tight-binding approximation is that the projected plane waves form an over-complete set. A well-defined basis for a crystal with $N_{\mathrm{s/c}}$ sites per unit cell consist of a subset with momenta in $N_{\mathrm{s/c}}$ Brillouin zones. However, there are different types of Brillouin zones depending on the phase factor $e^{i\boldsymbol{G} \cdot \boldsymbol{t}}$, where $\boldsymbol{t}$ are the relative positions of the sites inside the unit cell \cite{jung_imaging_2010}. For instance, in the honeycomb lattice there are 3 types of BZ, since $e^{-i\boldsymbol{G}\cdot\boldsymbol{t}} = e^{ia2\pi/3}$, {with} $a \in \mathbb{Z}_3$ (see Supplemental Material (SM)~\cite{SuppMat} \ref{app:spillage_TB}). This issue can be handled by replacing the sum over reciprocal lattice vectors $\boldsymbol{G}$ by an average over the different types of $\boldsymbol{G}$, and multiplying by $N_{\mathrm{s/c}}$. 

With these modifications, the structural spillage Eq.~\eqref{eq:qB-spillage} can be defined in the tight-binding approximation as
\begin{equation}
    \gamma_{\mathrm{qB}}^{\mathrm{TB}}(\boldsymbol{k}) = \frac{1}{2} \frac{N_{\mathrm{s/c}}}{N_{\mathrm{BZs}}} \sum_{\boldsymbol{G}\in\mathrm{BZs}} \text{tr} \left[ \left( P_{\boldsymbol{k}+\boldsymbol{G}} - \tilde{P}_{\boldsymbol{k}+\boldsymbol{G}} \right)^2 \right],
\label{eq:qB-spillage-no-scatt-matrix-TB}
\end{equation}
where the sum over $\boldsymbol{G}$ runs over one BZ of each of the $N_{\mathrm{BZs}}$ types, the trace acts over the {internal} degrees of freedom $\alpha$, and we have defined the single-momentum projector $P_{\boldsymbol{p}}^{\alpha\beta} = P_{\boldsymbol{p},\boldsymbol{p}}^{\alpha\beta}$. 

Eqs.~\eqref{eq:qB-spillage-no-scatt-matrix-TB} and ~\eqref{eq:qB-spillage} define the structural spillage to be used in the tight-binding approximation and first-principles calculations, respectively. In the remainder of the paper, we demonstrate how they capture topological phase transitions of amorphous systems, using low-dimensional bismuth as an example.

\paragraph*{Tight-binding benchmark: bismuthene on a substrate-.}
Crystalline bismuthene consists of a 2D honeycomb monolayer of bismuth atoms. Experiments suggest it to be a quantum spin Hall insulator with topological helical edge states when grown on SiC(0001)~\cite{reis_bismuthene_2017} or Ag(111)~\cite{sun_epitaxial_2022} substrates. The effect of the substrate is crucial: it filters the $p_z$ orbitals away from the Fermi level leaving the $p_{x,y}$ orbitals, resulting in a large gap ($\sim 0.67\mathrm{eV}$) and a non-zero strong $\mathbb{Z}_2$ topological index. Moreover, amorphous bismuthene on a substrate is predicted to remain topological via first-principles calculations \cite{costa_toward_2019,focassio_structural_2021}, making it a convenient system to benchmark our proposed structural spillage.

The low-energy physics of bismuthene is captured by a tight-binding model with $p_{x,y}$ orbitals in the honeycomb lattice, coupled by nearest-neighbour hoppings $t_{\sigma}$ and $t_{\pi}$, a large onsite SOC $\lambda$, and a substrate-induced Rashba SOC $\lambda_R$ (which we take proportional to $\lambda$)~\cite{reis_bismuthene_2017}. To extend this model to amorphous structures while preserving the short-range order expected in amorphous systems \cite{Zallen}, we use the voronization of a pointset \cite{Mitchell2018,Marsal2020} (see SM~\cite{SuppMat} \ref{app:bismuthene}). When the pointset is triangular, the voronization produces its dual honeycomb lattice. By randomly displacing the triangular pointset according to a characteristic length $r$, the voronization produces lattices with threefold coordination, as the honeycomb lattice, but with a finite density of non-hexagonal plaquettes (see Fig.~\ref{fig:fig_panel}(a))~\cite{Grushin2022}. Therefore, $r$ continuously controls how amorphous are our lattices, allowing us to study the effect of structural disorder on topological properties. In the following, we quantify how amorphous our systems are by the (configuration-averaged) density of non-hexagonal plaquettes $\rho_{\text{non-hex}}$, which is in one-to-one correspondence to the parameter $r$ (see SM~\cite{SuppMat} \ref{app:bismuthene}).

In Fig.~\ref{fig:fig_panel} we present the topological phase diagram of amorphous bismuthene as a function of $\rho_{\text{non-hex}}$ and $\lambda$, benchmarking $\gamma_{\mathrm{qB}}^{\mathrm{TB}}(\boldsymbol{k})$ against the two-terminal conductance results. In the crystalline limit ($\rho_{\text{non-hex}}=0$), the system starts as a Dirac semimetal for vanishing $\lambda$, and a finite $\lambda$ opens up a topological gap, similarly to graphene~\cite{kane_z_2_2005}. Above a critical $\lambda$, where the gap closes at the $\Gamma$ point, the system becomes a topologically trivial insulator, adiabatically connected to the atomic limit in which only the onsite SOC is non-zero. 

Both the conductance (Fig.~\ref{fig:fig_panel}(c)) and the structural quasi-Bloch spillage (Fig.~\ref{fig:fig_panel}(d)) capture the topological transition, even at finite structural disorder ($\rho_{\text{non-hex}} \neq 0$). The conductance in the topological insulator phase is equal to $2e^2/h$, originating from the helical edge states, while it reduces to zero after the phase transition to the trivial insulator. Concomitantly, $\gamma_{\mathrm{qB}}^{\mathrm{TB}}(\boldsymbol{k}=0)$ is large in the topological phase and small in the trivial phase because we choose the reference system to be a trivial crystal, only with non-zero onsite $\lambda$. Had we chosen the topological state as reference, the magnitude of the spillage in each phase would be inverted; see SM~\cite{SuppMat} \ref{app:bismuthene}. The critical $\lambda$ at the transition for the crystal is correctly predicted by $\gamma_{\mathrm{qB}}^{\mathrm{TB}}(\boldsymbol{k}=0)$. In agreement with Refs. \cite{costa_toward_2019,focassio_structural_2021}, we find that increasing disorder decreases the topological gap and hence the critical $\lambda$. Nevertheless, the realistic value of $\lambda \simeq 0.22 t_{\sigma}$~\cite{reis_bismuthene_2017} lies in the topological phase also in the amorphous case.

Lastly, Fig.~\ref{fig:fig_panel}(b) shows $\gamma_{\mathrm{qB}}^{\mathrm{TB}}(\boldsymbol{k})$ for fixed $\lambda = 0.22 t_{\sigma}$ and $\rho_{\text{non-hex}}=0.53$. $\gamma_{\mathrm{qB}}^{\mathrm{TB}}(\boldsymbol{k})$ is peaked around $\boldsymbol{k} = 0$ with a value $\sim 1.5$, reminiscent of the crystalline topological band inversion occurring at the $\Gamma$ point. 

\begin{figure}
    \centering
    \includegraphics[width = \columnwidth]{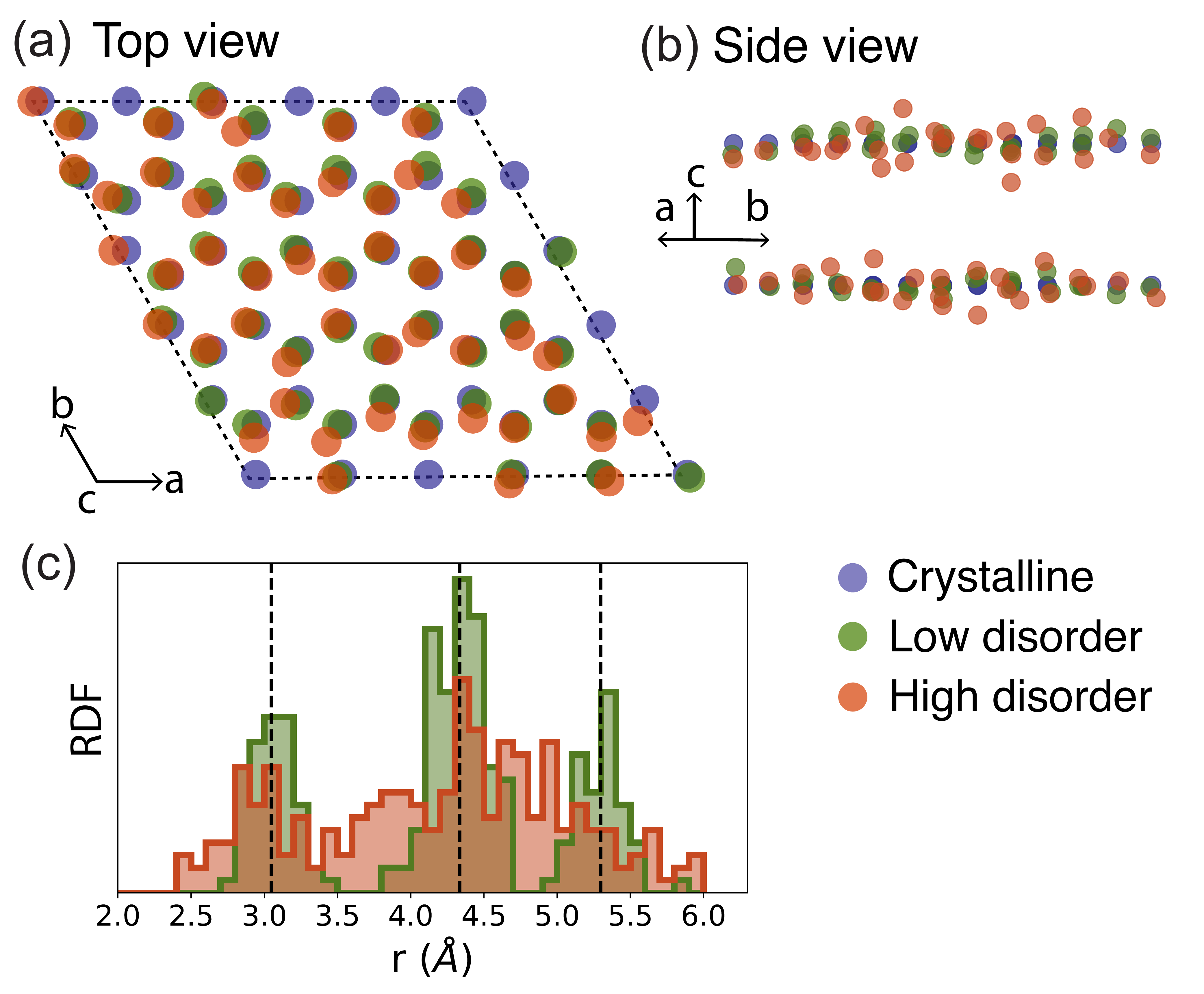}
    \caption{Bismuth bilayer supercells used in DFT calculations. 
    (a) and (b) show in-plane and out of plane views of the supercell, respectively. 
    The colors indicate different degrees of disorder: crystal (blue), low disorder (green) and high-disorder (orange).
    (c) Radial distribution function (RDF) showing the statistics of the bond lengths in the disordered bismuth bilayer and their deviations from the perfect crystal (vertical dashed lines). 
    The disorder is sampled from a Gaussian distribution with a standard deviation of $0.15~\angstrom$ for the low disorder and $0.30~\angstrom$ for the high disorder.}
    \label{fig:DFTstructure}
\end{figure}

\paragraph*{Structural spillage in DFT: free-standing Bi bilayer-.} 
To show that Eq.~\eqref{eq:qB-spillage} is well suited for high-throughput screening of amorphous topological materials, we calculate the structural spillage from the output wavefunctions of first-principles calculations (see full details in SM). We choose previously-studied free-standing bismuth (111) bilayer as an example. This 2D bismuth allotrope, whose crystalline phase consists of a buckled honeycomb lattice with lattice constant $a=4.33~\angstrom $, is also predicted to be a strong topological insulator crystal with $\mathbb{Z}_2 = 1$ \cite{murakami_quantum_2006,wada_localized_2011,liu_stable_2011,huang_nontrivial_2013}. However, no prediction exists for its amorphous counterpart. 

To represent amorphous structures given the periodic boundary conditions of the calculations, we create $5 \times 5 \times 1$ supercells comprising of 50 Bi atoms per bilayer. Their electronic structure is calculated for a single supercell momentum, the center of the supercell BZ. Starting from a crystalline supercell, the structure is disordered by adding random displacements in the $x,y,$ and $z$ directions, sampled from a Gaussian distribution. The structures and their corresponding radial distribution functions are shown in Fig.~\ref{fig:DFTstructure}.

\begin{figure}
    \centering
    \includegraphics[width = \columnwidth]{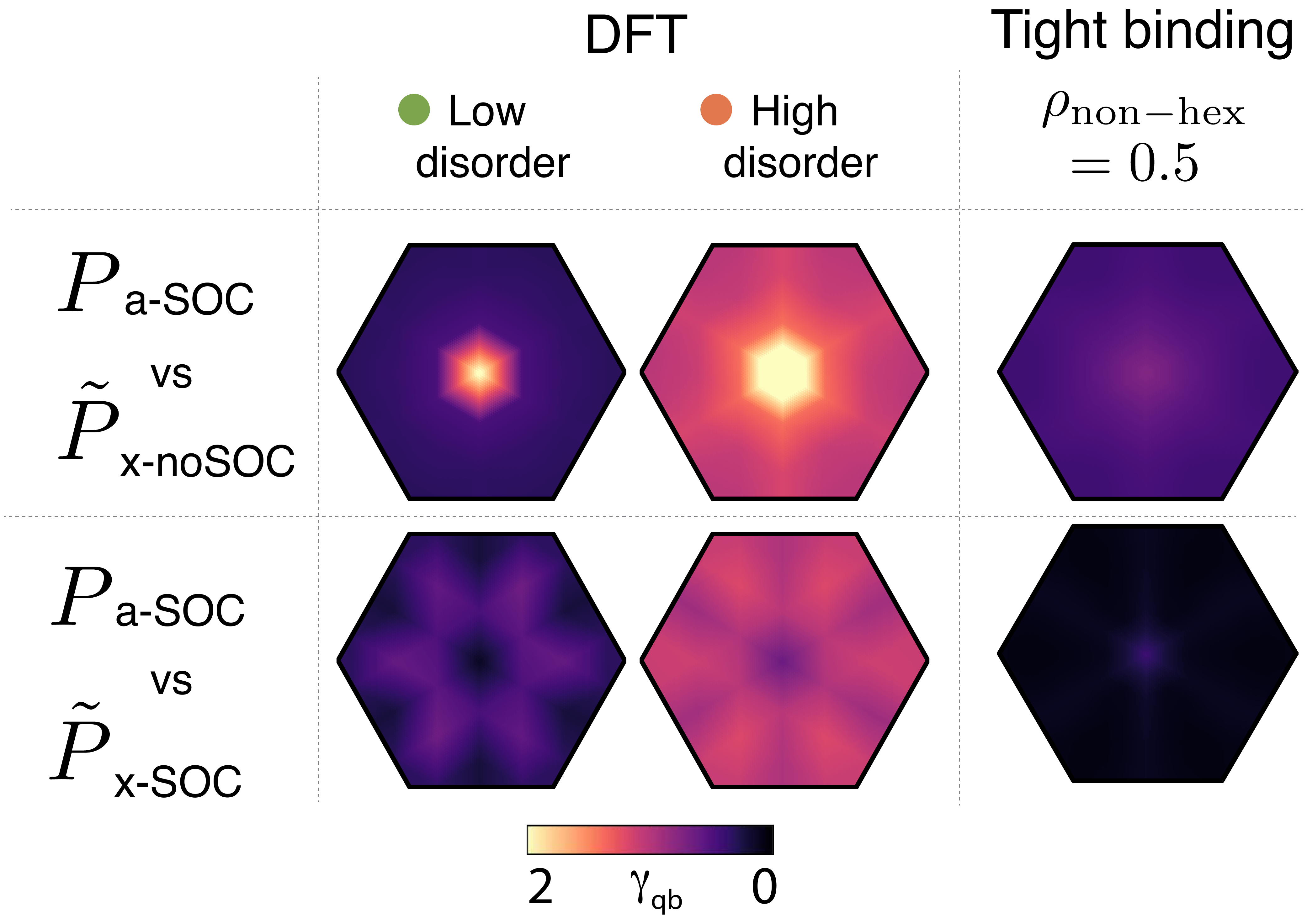}
    \caption{Structural quasi-Bloch spillage $\gamma_{\mathrm{qB}}(\boldsymbol{k})$ for the bismuth bilayer. First row: comparison between an amorphous system with SOC (a-SOC) and a crystalline system without SOC (x-noSOC). Comparing an amorphous system without SOC with a crystalline sample with SOC leads to similar results. 
    Second row: comparison between the amorphous and crystalline systems with SOC (a-SOC and x-SOC, respectively). 
    $\gamma_{\mathrm{qB}}(\boldsymbol{k})$ is high at $\boldsymbol{k} = 0$ for the first row while small for the second row, indicating that amorphous bismuth bilayer is a topological insulator.
    The last column shows a comparison with the tight-binding quasi-Bloch spillage $\gamma_{\mathrm{qB}}^{\mathrm{TB}}(\boldsymbol{k})$ (see SM~\cite{SuppMat} \ref{app:bilayer}).}
    \label{fig:DFTspillage}
\end{figure}

To predict the topological phase of amorphous Bi bilayer with SOC we compute Eq.~\eqref{eq:qB-spillage} with plane-wave-based DFT (see SM~\cite{SuppMat} \ref{app:DFT}) to compare it with its crystalline counterpart without and with SOC. When SOC is not included, and hence when it is topologically trivial (Fig.~\ref{fig:DFTspillage}, first row), $\gamma_{\mathrm{qB}}(\boldsymbol{k})$ is peaked at $\boldsymbol{k} = 0$, with $\gamma_{\mathrm{qB}}(\boldsymbol{k}=0)>2$. Increasing disorder smooths $\gamma_{\mathrm{qB}}(\boldsymbol{k})$, yet it remains peaked at $\Gamma$ with a value greater than 2. In contrast, when we include SOC in calculations of both the disordered Bi bilayer and the pristine crystal (Fig.~\ref{fig:DFTspillage}, second row) the spillage is always small. Both rows together show that amorphous bismuth bilayer with SOC is in the same topological state as the crystal with SOC, a strong topological insulator crystal with $\mathbb{Z}_2 = 1$.

We have performed a similar analysis using a tight-binding model for the amorphous Bi (111) bilayer (introduced in SM~\cite{SuppMat} \ref{app:bilayer}). The results, displayed in the last column of Fig.~\ref{fig:DFTspillage}, show that for comparable disorder strengths $\gamma_{\mathrm{qB}}^{\mathrm{TB}}(\boldsymbol{k})$ is broader and its maximum value is smaller than $\gamma_{\mathrm{qB}}(\boldsymbol{k})$ in DFT. It is thus apparent that, due to the approximations in the tight-binding calculation of the spillage, which lacks information of the real space extension of the orbitals, the spillage method is more suitable for DFT, an advantageous feature compared to other topological indicators available for 
non-crystalline systems.

\paragraph*{Discussion-.} 
We have introduced the structural spillage as an efficient method to signal non-crystalline topological phases, compatible with tight-binding and \textit{ab-initio} simulations. We have used it to predict amorphous Bi bilayer as a novel topological insulator.

As was the case for spin-orbit spillage in crystals, we expect the structural spillage to signal a large fraction of promising materials, but not to be infallible: if multiple band inversions are introduced upon amorphization, the spillage might also be artificially large. However, unlike for crystals, the spillage is currently the only systematic, model-independent method that is compatible with \textit{ab-initio} calculations. Additionally, we observe that, for different disorder realizations, its fluctuations are smaller compared to scattering methods like calculating the conductance. It can also be applied to systems without a spectral gap, where the effective Hamiltonian approach~\cite{Varjas2019} can fail~\cite{Marsal2020}. Lastly, while Eq.~\eqref{eq:qB-spillage} is general, the definition of the spillage is relatively versatile and can accommodate less standard cases. For example, when no crystalline counterpart exists, one may define a plane-wave-resolved spillage (see SM~\cite{SuppMat} \ref{app:pw-spillage}) by using Eq.~\eqref{eq:qB-spillage-general} without the sum over $\boldsymbol{G}$, a modification worth studying in the future.

The structural spillage establishes a clear road-map to construct a high-throughput catalogue of non-crystalline (amorphous, polycrystalline, quasicrystalline) topological materials by screening existing amorphous databases, or by scrutinizing realistic structures obtained using existing \textit{ab-initio} molecular dynamics packages~\cite{Kuhne2020}. This methodology may enable for the first time the systematic prediction and discovery of a potentially large number of amorphous materials that are currently inaccessible, suitable to develop affordable and scalable topological devices.

\paragraph*{Acknowledgements-.} 
We are grateful to S. Franca, F. de Juan, J. Hannukainen, D. L\'{o}pez-Cano, R. Queiroz, Q. Marsal, A. Soluyanov, R. M. Martin, and J. Vinson for fruitful discussions and related collaborations. This work was partially funded by the U.S. Department of Energy, Ofﬁce of Science, Ofﬁce of Basic Energy Sciences, Materials Sciences and Engineering Division under Contract No. DE-AC02-05-CH11231 within the Nonequilibrium Magnetic Materials Program (MSMAG), specifically the work by P.C., F.H., and S.M.G. D.M.S. is supported by an FPU predoctoral contract from Spanish MCIU No. FPU19/03195. A.G.G. acknowledges financial support from the European Research Council (ERC) Consolidator grant under grant agreement No. 101042707 (TOPOMORPH).  D.V. was supported by the Swedish Research Council (VR) and the Knut and Alice Wallenberg Foundation. Computational resources were provided by the National Energy Research Scientific Computing Center and the Molecular Foundry, DOE Office of Science User Facilities supported by the Office of Science, U.S. Department of Energy under Contract No. DEAC02-05CH11231. The work performed at the Molecular Foundry was supported by the Office of Science, Office of Basic Energy Sciences, of the U.S. Department of Energy under the same contract.

\paragraph*{Author contributions-.} 
The original idea was conceived by P.C. and S.M.G. D.M.S. derived the expressions for the quasi-Bloch spillage and performed tight-binding calculations assisted by A.G.G.  P.C. performed the DFT calculations and developed the spillage code for plane-waves assisted by S.M.G., F.H. and D.V. All authors contributed to the interpretation of results and writing of the manuscript. A.G.G. and S.M.G. supervised the project.

\bibliographystyle{apsrev4-2}
\bibliography{bib_notes_amorphous_spillage}

\clearpage
\newpage

\clearpage
\newpage

\setcounter{secnumdepth}{5}
\renewcommand{\theparagraph}{\bf \thesubsubsection.\arabic{paragraph}}

\renewcommand{\thefigure}{S\arabic{figure}}
\setcounter{figure}{0}

\appendix

\onecolumngrid

\tableofcontents

\section{\label{app:TB}Tight-binding models}

This Appendix describes the method for generating the amorphous tight-binding models used in the maint text. We include as well further calculation details and some additional discussion regarding the phase diagrams that one can obtain using different reference systems of the structural spillage.

\subsection{\label{app:bismuthene}Model for bismuthene on a substrate}

This section describes how to generate the amorphous bismuthene structure and tight-binding Hamiltonian that we have used to benchmark the structural spillage method in Fig.~\ref{fig:fig_panel}.

\subsubsection{Tight-binding Hamiltonian}

Crystalline bismuthene consists of a 2D honeycomb monolayer of bismuth atoms \cite{reis_bismuthene_2017}. An effective tight-binding of crystalline bismuthene on a substrate was proposed by Ref.~\cite{reis_bismuthene_2017}. It consists of $p_x$ and $p_y$ orbitals in the honeycomb lattice, coupled by nearest-neighbour hoppings, a large onsite SOC, and a substrate-induced Rashba SOC. In real space and in the basis $\left\{p_{x\uparrow},p_{x\downarrow},p_{y\uparrow},p_{y\downarrow}\right\}$, the Hamiltonian reads:
\begin{equation}
\begin{split}
    H = & - \frac{1}{2} \sum_{\langle i j\rangle} \left[ \left( t_{\sigma} - t_{\pi} \right) \tau_0 + \left( t_{\sigma} + t_{\pi} \right) \left( c_{ij}^{(2)} \tau_z + s_{ij}^{(2)} \tau_x \right) \right] \sigma_0 + \sum_i \left[ \lambda \tau_y \sigma_z \right] + \\
    & + \sum_{\langle i j\rangle} i \left\{ \lambda_R^A \tau_0 \left[ s_{ij} \sigma_x - c_{ij} \sigma_y \right] + \lambda_R^E \left[ \left( c_{ij} \tau_x - s_{ij} \tau_z \right) \sigma_x - \left( c_{ij} \tau_z + s_{ij} \tau_x \right) \sigma_y \right] \right\},
\end{split}
\label{eq:H_bismuthene}
\end{equation}
where we have defined $c_{ij} = \cos(\theta_{ij})$, $s_{ij} = \sin(\theta_{ij})$, $c_{ij}^{(2)} = \cos(2\theta_{ij})$, and $s_{ij}^{(2)} = \sin(2\theta_{ij})$, with $\theta_{ij}$ the angle between the bond joining site $i$ to site $j$ and the $x$ axis. $\tau_{\mu}$ and $\sigma_{\mu}$ are the Pauli matrices acting on the orbital $\left\{p_{x},p_{y}\right\}$ and spin $\left\{\uparrow,\downarrow\right\}$ degrees of freedom, respectively. $t_{\sigma}$ and $t_{\pi}$ are the sigma and pi nearest-neighbour hoppings, $\lambda$ is the onsite SOC, and $\lambda_R^A$ and $\lambda_R^E$ are the orbital-independent and orbital-dependent Rashba SOC, respectively. As in Ref.~\cite{reis_bismuthene_2017}, in this work we will assume that $\lambda_R^A = \lambda_R^E = \lambda_R$. The values used in Ref.~\cite{reis_bismuthene_2017} are $t_{\sigma} \simeq 2.0 \mathrm{eV}$, $t_{\pi} \simeq 0.21 \mathrm{eV} \simeq 0.11 t_{\sigma}$, $\lambda \simeq 0.44 \mathrm{eV} \simeq 0.22 t_{\sigma}$, and $\lambda_R \simeq 0.032 \mathrm{eV} \simeq 0.074 \lambda$. In our calculations, we will take $t_{\sigma}$ as the unit of energy, we will use the same value for $t_{\pi} = 0.11 t_{\sigma}$, and we will vary both the onsite SOC $\lambda$ as well as the Rashba SOC proportionally to the former, $\lambda_R = 0.074 \lambda$.

The Hamiltonian \eqref{eq:H_bismuthene} can readily be applied to an amorphous lattice once we define which sites are nearest neighbours of each other. In principle, it could be generalized to include a dependence on the distance in the hoppings, such as the Harrison law \cite{harrison_book_chemical_bond}. However, we will consider fixed values for the hoppings, which can be a good approximation for covalently-bonded amorphous solids, which usually display a rather narrow distribution of bond distances \cite{Zallen}. Moreover, this approximation enables us to isolate the effect of structural disorder.

\subsubsection{\label{app:amorphous_bismuthene}Construction of amorphous structures}

Covalently-bonded amorphous materials usually preserve local environments similar to the ones in the corresponding crystals, since they are set by the strong covalent bonds. Therefore, most amorphous materials have average coordination numbers, bond distances, bond angles, etc., which are centered around those of the crystal~\cite{Zallen}. With this in mind, our amorphous models preserve, for every site, the threefold coordination of the honeycomb lattice. This is achieved by applying the Voronoi method similar to Ref.~\cite{Marsal2020}, but with a modification that enables us to control the degree of amorphization. 

In particular, we first construct a pointset forming a triangular lattice with lattice constant $a$, whose points will be called seeds. We then randomly displace the seeds from their initial positions following an exponential distribution with characteristic distance $r \cdot a$ in the radial direction, and a uniform distribution in the angular direction. We thereafter compute their corresponding Voronoi diagram, which is defined by the Voronoi cells, i.e., the regions consisting of all points closer to one seed point than to any other. The vertices of such cells, called Voronoi vertices, form a threefold coordinated lattice with the edges of the Voronoi cells corresponding to the nearest-neighbour bonds (only the vertices at the boundaries of the system have fewer than three neighbours). 

\begin{figure}
    \centering
    \includegraphics[width=\textwidth]{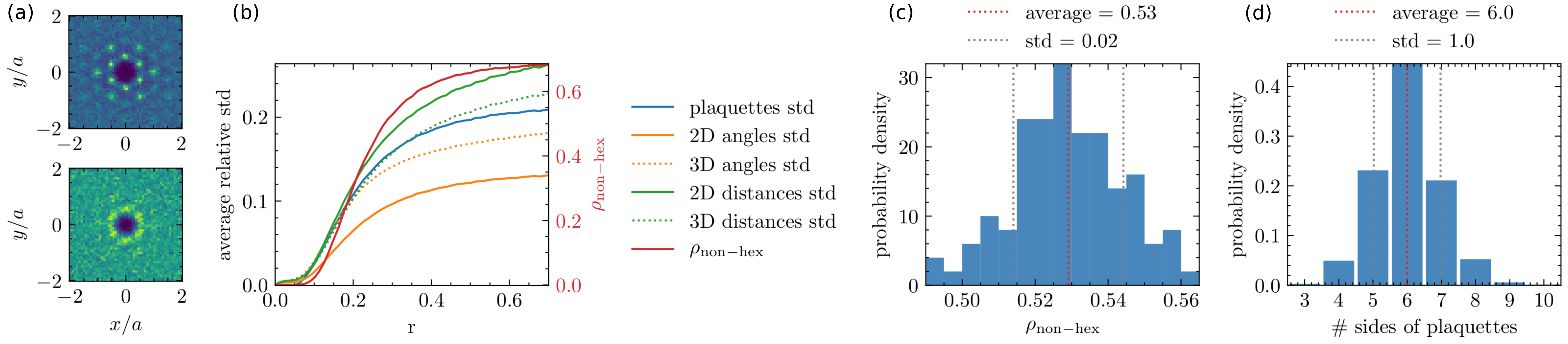}
    \caption{(a) Histograms of the relative positions of atoms for two amorphous structures with different disorder strengths, $r=0.3$ (top) and $r=0.5$ (bottom). 
    (b) Configuration-averaged structural quantities as a function of the parameter $r$ controlling the amorphousness: standard deviations (std) of the distributions of nearest neighbour bond angles, bond distances (both for the planar bismuthene as well as for the buckled Bi bilayer), and plaquettes, as well as density of non-hexagonal plaquettes. For each disorder intensity $r$, the results have been averaged over 100 different realizations. 
    (c) Distribution of the ratios of non-hexagonal plaquettes $\rho_{\text{non-hex}}$ obtained with $100$ disorder realizations with fixed disorder $r=0.3$. 
    (d) Distribution of plaquettes for a given disorder realization with $r=0.3$ (corresponding to $\rho_{\text{non-hex}} \simeq 0.55$).}
    \label{Sfig:structural_statistics}
\end{figure}

The lattices obtained in this way have large variances in the bond angle and bond length distributions, which might not be very realistic. In order to reduce this artifact, we apply a simple iterative relaxation procedure. We select the threefold coordinated sites one by one and displace them to the barycenter formed by their three nearest neighbours. We iterate this process until convergence is reached, i.e., until the displacements are smaller than some small cutoff. This relaxation procedure tends to set the bond angles as close as possible to the crystalline angle, $120^{\circ}$. Finally, once the lattice is relaxed, we rescale the distances so that the average nearest-neighbour distance is $a/\sqrt{3}$, which is the corresponding value in the crystalline honeycomb lattice. Fig.~\ref{Sfig:structural_statistics}(a) shows the resulting histograms of the relative positions of atoms for two amorphous structures with different disorder strengths, $r=0.3$ (top) and $r=0.5$ (bottom). Both structures are isotropic at long distances, although for small disorder the nanocrystalline domains (see for example Fig.~\ref{fig:fig_panel}(a) in the main text) give rise to broad nearest neighbour peaks around the crystalline positions. For high disorder, the correlation hole for distances under $a/\sqrt{3}$ and an annular peak are visible.

The parameter $r$, characterizing the exponential distribution by which the seeds are displaced from the regular triangular lattice, continuously controls the amorphousness of the resulting Voronoi lattice. Indeed, since the Voronoi diagram of a triangular lattice is a honeycomb lattice, we recover the crystal in the $r \rightarrow 0$ limit. Increasing $r$ introduces non-hexagonal plaquettes in the Voronoi lattice, at least until $r \gtrsim 1$, when the seed becomes completely random (since all the information from the initial triangular seed is lost). This can be observed in Fig.~\ref{Sfig:structural_statistics}(b), which shows that the configuration-averaged standard deviations of the distributions of bond angles, bond distances, and plaquettes start to saturate at about $r \gtrsim 0.6$. 

Structural disorder can be quantified by several properties. These include the standard deviations of the distributions of nearest-neighbour distances, angles and plaquettes (normalized by the corresponding average values), as well as the density of non-crystalline plaquettes (in our models, where the crystalline limit consists of a honeycomb lattice, the non-crystalline plaquettes correspond to the non-hexagonal ones). In order to take into account the finite-size effects, for each parameter $r$, we consider the configuration-average of these quantities over 100 realizations. 

As shown in Fig.~\ref{Sfig:structural_statistics}(b), all these configuration-averaged quantities have the same qualitative dependence with the parameter $r$. In particular, there exists a one-to-one correspondence between our control parameter $r$ and any of these configuration-averaged quantities. However, for particular disorder realizations in a finite system, there are fluctuations that make their relation to $r$ not one-to-one before performing the configuration average. This is illustrated by the distribution of ratios of non-hexagonal plaquettes $\rho_{\text{non-hex}}$ shown in Fig.~\ref{Sfig:structural_statistics}(c) for different realizations with fixed $r=0.3$. Therefore, we have chosen to physically characterize the amorphousness of a system by the configuration-averaged density of non-hexagonal plaquettes formed by the nearest neighbour sites $\rho_{\text{non-hex}}$. This measure could be generalized to other models whose crystalline limit consisted of lattices other than the honeycomb. Finally, Fig.~\ref{Sfig:structural_statistics}(d) shows an example distribution of plaquettes obtained for a particular disorder realization with $r=0.3$, which corresponds to $\rho_{\text{non-hex}} \simeq 0.55$, while the configuration-average for this $r$ corresponds to $\rho_{\text{non-hex}} \simeq 0.53$.

The above procedure generates structures with open boundary conditions, which is useful to compute e.g. the local density of states at the edges or the longitudinal conductance once some leads have been attached. However, for spectral quantities such as the spillage, we can reduce the possible finite-size effects by imposing periodic boundary conditions, or equivalently by putting the system on a torus. An amorphous system might have a different number of atoms at opposite edges, so the periodic boundary conditions cannot be imposed directly, but rather before computing the Voronoi tessellation, as described below. 

Before explaining the procedure to impose the periodic boundary conditions, let us note that our periodic systems consist of a rectangular supercell with sides $L_x$ and $L_y$. In order for the periodic boundary conditions to be applicable to systems with an arbitrary amount of structural disorder, including the crystalline limit, $L_x$ and $L_y$ are restricted to the values such that the supercell is commensurate with the initial crystalline unit cell. In our models, where the crystalline limit is a honeycomb lattice, the previous condition imposes that $L_x = n_x a$ and $L_y = n_y \sqrt{3} a$, where $a$ is the lattice constant, and $n_x,n_y$ are integer numbers.

Taking this into account, let us now describe the procedure to impose periodic boundary conditions on a system with an arbitrary amount of disorder. First, we generate a triangular seed within the supercell $x\in\left[0,L_x\right)$, $y\in\left[0,L_y\right)$, and we disorder choosing a finite value of $r$. Then, we repeat this initial seed in the eight nearest-neighbour supercells, i.e., we copy the seed points displaced from their initial positions $\boldsymbol{x}$ to $\boldsymbol{x} + \boldsymbol{L} = \boldsymbol{x} +  \left(n_x L_x, n_y L_y\right)$, with $n_x,n_y \in \left\{1,0,-1\right\}$. Then, the Voronoi tessellation of the whole system (composed by the nine supercells) is determined. This gives rise to a threefold coordinated lattice with the following convenient feature: the supercell defined by the sites inside the region $x\in\left[0,L_x\right)$, $y\in\left[0,L_y\right)$ has the same number of sites in opposite sides. Therefore, the periodic boundary conditions can be now applied to this supercell (all the sites outside this supercell are discarded). Finally, we carry out the relaxation procedure of this supercell, being careful to preserve the periodic boundary conditions. 

To conclude this section, let us mention that we generate the systems with open boundary conditions starting from a system with periodic boundary conditions, by first removing the bonds at the edges of the supercell and then removing the dangling sites. This way, the bulk of the periodic structure where the spillage is computed is the same as the bulk of the open system where the conductance is determined, which allows us to safely compare their predictions of the topological phase. 

\subsubsection{Additional results: density of states and structural spillage for different reference systems}

\begin{figure}[!t]
    \centering
    \includegraphics[width=\textwidth]{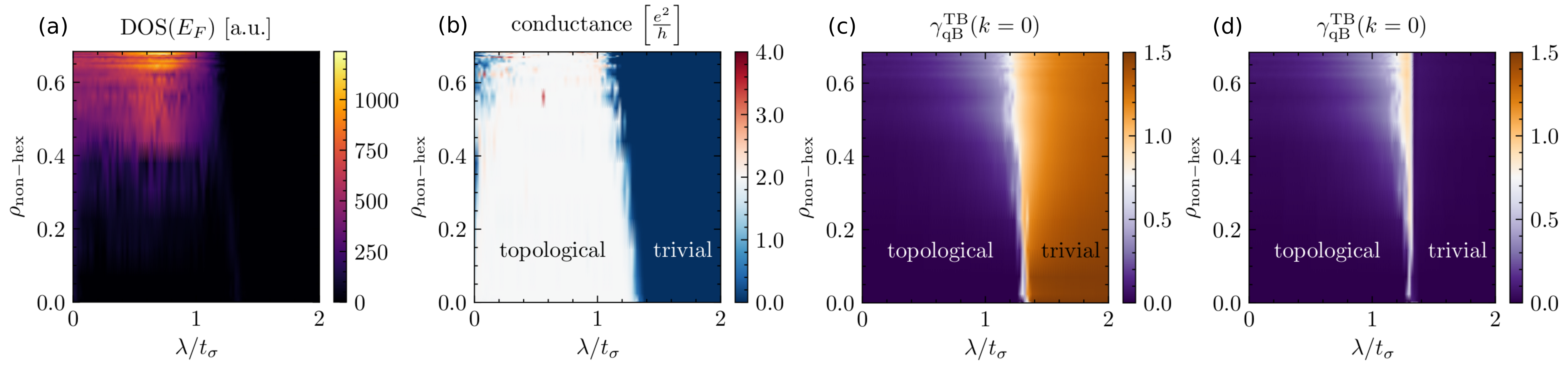}
    \caption{Phase diagrams of different quantities as a function of SOC $\lambda$ and amorphousness $\rho_{\text{non-hex}}$ for the bismuthene model. (a) Density of states at the Fermi level of the system with periodic boundary conditions. (b) Two-terminal longitudinal conductance in the ``zigzag'' ribbon. (c) Structural quasi-Bloch spillage $\gamma_{\mathrm{qB}}^{\mathrm{TB}}(\boldsymbol{k}=0)$ comparing the amorphous system with a topological bismuthene crystal with $\lambda=0.1t_{\sigma}$. (d) Structural quasi-Bloch spillage $\gamma_{\mathrm{qB}}^{\mathrm{TB}}(\boldsymbol{k}=0)$ comparing the amorphous system with SOC $\lambda$ to the corresponding crystal with the same SOC $\lambda$.}
    \label{Sfig:additional_results_bismuthene}
\end{figure}

In this section we discuss further different phase diagrams that may be obtained for the bismuthene tight-binding model and its spillage in the tight-binding approximation. Fig.~\ref{Sfig:additional_results_bismuthene} shows phase diagrams for the density of states, conductance and structural spillage corresponding to the same bismuthene structures as the ones presented in the main text in Fig.~\ref{fig:fig_panel}. In particular, Fig.~\ref{Sfig:additional_results_bismuthene}(a) shows that the density of states at the Fermi level increases with $\rho_{\text{non-hex}}$ when the SOC is such that the crystal is in the topological phase ($\lambda \lesssim 1.3 t_{\sigma}$). This is due to the band broadening due to the disorder, and also from the appearance of low-energy states induced by a sublattice imbalance in a bipartite lattice \cite{lieb_two_1989}. At high disorder, this induces the band inversion that drives the system from topological to trivial at a smaller SOC than in the crystal. 

In order to show that the quantized conductance does not arise from disorder-robust trivial edge states present in one particular crystalline direction, we display in Fig.~\ref{Sfig:additional_results_bismuthene}(b) the longitudinal two-terminal conductance along the direction perpendicular to the one displayed in the main Fig.~\ref{fig:fig_panel} (the edges here would correspond to a zigzag ribbon in the crystalline case). As expected, both conductances coincide, which is a signature of the topological helical edge states, which live at all the boundaries of the system.

Let us now explore how the structural spillage changes when we choose a topological reference system, as opposed to a trivial reference system used in the main text, Fig.~\ref{fig:fig_panel}. Fig.~\ref{Sfig:additional_results_bismuthene}(c) shows the structural quasi-Bloch spillage when the reference system is a topological crystal with SOC $\lambda = 0.1 t_{\sigma}$. Contrary to the trivial reference case shown in the main in Fig.~\ref{fig:fig_panel}, now the spillage is small in the topological phase and large in the trivial one, as expected from Fig.~\ref{fig:fig1}. Importantly, the transition is predicted at approximately the same SOC irrespective of the reference system, which shows the robustness of the spillage. 

Finally, in order to isolate the effect of the structural disorder on the topological band inversion from the effect of SOC, we have also computed the structural quasi-Bloch spillage comparing each amorphous system with amorphousness $\rho_{\text{non-hex}}$ and SOC $\lambda$ to a reference crystal with the same SOC $\lambda$, shown in Fig.~\ref{Sfig:additional_results_bismuthene}(d). This choice highlights the regions where disorder induces a topological band inversion. For example, if the reference crystal is topological for a given $\lambda$, this spillage will have a large value if the disorder induces a trivial state. Therefore, interpreting Fig.~\ref{Sfig:additional_results_bismuthene}(d) requires knowledge of the topological phase of the crystal at each $\lambda$. For $\lambda \lesssim 1.3 t_{\sigma}$, the reference crystal is topological. Since the spillage is small for $\lambda \lesssim 1.1 t_{\sigma}$, the amorphous system is topological for $\lambda \lesssim 1.1 t_{\sigma}$. However, at high disorder, the spillage becomes large between $\lambda \simeq 1.1 t_{\sigma}$ and $\lambda \simeq 1.3 t_{\sigma}$, which indicates that the disorder induces a trivial phase. Lastly, for $\lambda \gtrsim 1.3 t_{\sigma}$, the reference crystal is trivial, and the spillage is low, indicating that the amorphous system is also trivial. 

In conclusion, all phase diagrams Fig.~\ref{Sfig:additional_results_bismuthene} (b-d) agree qualitatively. The spillage is able to predict the topological phase transition independent of the reference system.

\subsection{\label{app:bilayer}Model for free-standing bismuth (111) bilayer}

In this section, we introduce a tight-binding model for the amorphous bismuth bilayer, for which we  study the structural spillage. After introducing the model and describing the method to generate the amorphous structures, we analyze its topological phase diagram to further benchmark the structural spillage. Finally, we compare the tight-binding results and DFT calculations, as shown in Fig.~\ref{fig:DFTspillage}. We conclude that, while both qualitatively agree, the structural spillage method works better in DFT.

\subsubsection{Tight-binding Hamiltonian\label{app:bilayer_a}}

Crystalline bismuth (111) bilayer consists of a buckled honeycomb lattice of bismuth atoms, where each sublattice has a different height \cite{liu_stable_2011}. An effective tight-binding of crystalline Bi bilayer was introduced by Ref.~\cite{li_localized_2021}, where the three $p$ orbitals are relevant due to the absence of the substrate in this case. Their model consists of spinful $p_x$, $p_y$ and $p_z$ orbitals in the buckled honeycomb lattice with up to third nearest-neighbour hoppings. For simplicity, we will restrict ourselves to nearest-neighbour hoppings and onsite SOC. In real space and in the basis $\left\{p_{x\uparrow},p_{x\downarrow},p_{y\uparrow},p_{y\downarrow},p_{z\uparrow},p_{z\downarrow}\right\}$, the Hamiltonian reads:
\begin{equation}
\begin{split}
    H = & \sum_{\langle i j\rangle} \left[ t_{\pi} \tau_0 \sigma_0 - \left( t_{\sigma} + t_{\pi} \right) \begin{pmatrix} \left(\boldsymbol{d}_{ij}\cdot \boldsymbol{u}_x\right)^2 & \left(\boldsymbol{d}_{ij}\cdot \boldsymbol{u}_x\right) \left(\boldsymbol{d}_{ij}\cdot \boldsymbol{u}_y\right) & \left(\boldsymbol{d}_{ij}\cdot \boldsymbol{u}_x\right) \left(\boldsymbol{d}_{ij}\cdot \boldsymbol{u}_z\right) \\ \left(\boldsymbol{d}_{ij}\cdot \boldsymbol{u}_y\right) \left(\boldsymbol{d}_{ij}\cdot \boldsymbol{u}_x\right) & \left(\boldsymbol{d}_{ij}\cdot \boldsymbol{u}_y\right)^2 & \left(\boldsymbol{d}_{ij}\cdot \boldsymbol{u}_y\right) \left(\boldsymbol{d}_{ij}\cdot \boldsymbol{u}_z\right) \\ \left(\boldsymbol{d}_{ij}\cdot \boldsymbol{u}_z\right) \left(\boldsymbol{d}_{ij}\cdot \boldsymbol{u}_x\right) & \left(\boldsymbol{d}_{ij}\cdot \boldsymbol{u}_z\right) \left(\boldsymbol{d}_{ij}\cdot \boldsymbol{u}_y\right) & \left(\boldsymbol{d}_{ij}\cdot \boldsymbol{u}_z\right)^2 \end{pmatrix} \sigma_0 \right] + \\
    & + \sum_i \left[ E_{0z} \begin{pmatrix} 0 & 0 & 0 \\ 0 & 0 & 0 \\ 0 & 0 & 1 \end{pmatrix} \sigma_0 + \lambda \boldsymbol{L} \cdot \boldsymbol{\sigma} \right],
\end{split}
\label{eq:H_Bi_bilayer}
\end{equation}
where $E_{0z}$ is the difference between the onsite energy of the $p_z$ and $p_{x,y}$ orbitals, $\boldsymbol{d}_{ij}$ is the unit vector along the bond from site $i$ to site $j$, and $\boldsymbol{u}_a$, $a=x,y,z$, are the unit vectors along the three cartesian axes. We have also defined the angular momentum matrices $L_a$, which act on the orbital subspace $\left\{p_{x},p_{y},p_{z}\right\}$:
\begin{equation}
    L_x = \begin{pmatrix}  0 &  0 &  0 \\
                           0 &  0 & -i \\
                           0 &  i &  0 
          \end{pmatrix} 
    \hspace{20pt} ; \hspace{20pt} 
    L_y = \begin{pmatrix}  0 &  0 &  i \\
                           0 &  0 &  0 \\
                          -i &  0 &  0 
          \end{pmatrix} 
    \hspace{20pt} ; \hspace{20pt} 
    L_z = \begin{pmatrix}  0 & -i &  0 \\
                           i &  0 &  0 \\
                           0 &  0 &  0 
          \end{pmatrix}.
\end{equation}
In our calculations, we will take $t_{\sigma}$ as the unit of energy, and fix the value of $t_{\pi} = 0.25 t_{\sigma}$ and $E_{0z} = -0.4 t_{\sigma}$. We vary the onsite SOC $\lambda$. From the DFT-derived tight-binding model of Ref.~\cite{li_localized_2021}, we can estimate that the actual SOC for the Bi bilayer is $\lambda \sim 0.7 t_{\sigma}$. The height of the bilayer enters via the vectors $\boldsymbol{d}_{ij}$. Different DFT calculations have predicted heights ranging from $d_z = 0.35a$ to $d_z = 0.40a$ \cite{liu_stable_2011,huang_nontrivial_2013,singh_low-energy_2019,li_localized_2021}. In this work, we will use $d_z = 0.9a/\sqrt{6} \simeq 0.37 a$.

\subsubsection{Construction of amorphous structures}

Our structures of amorphous Bi bilayers are constructed in a similar way to monolayer bismuthene. Indeed, the first step is generating an amorphous bismuthene lattice following the procedure outlined in Appendix \ref{app:amorphous_bismuthene}. We then have to assign different heights to the sites. In the crystalline limit, each sublattice has a different fixed height because of the buckling. Sublattices are no longer well-defined in an amorphous lattice, but we can still define some effective sublattices. One differentiating property between the two sublattices in a crystalline honeycomb lattice is the direction of their nearest-neighbour bonds: if the bonds from sublattice $A$ point at polar angles $\theta_{1}^A = \pi/2$, $\theta_{2}^A = -11\pi/12$ and $\theta_{3}^A = -\pi/12$, then the ones from sublattice $B$ point at $\theta_{1}^B = -\pi/2$, $\theta_{2}^B = \pi/12$ and $\theta_{3}^B = 11\pi/12$. Therefore, $\eta(S) = \mathrm{sign} \left[ \left( \sum_{l} \theta_l^S \text{ } \mathrm{mod} \text{ } 2\pi \right) - \pi \right]$ is equal to $+1$ for sublattice $S=A$ and $-1$ for $S=B$. Using $\eta(S) = \pm 1$ to define the effective sublattices in the amorphous structures, we then assign a height $\pm d_z/2$. Finally, we add some random disorder to the height of each site sampled from a Gaussian distribution with standard deviation $r_z \cdot a$. In particular, we choose the height disorder $r_z$ proportional to $r$, the parameter that controls the in-plane amorphousness. In the calculations presented in this work, we take $r_z = r d_z/(4a) \simeq 0.09 r$. Fig.~\ref{Sfig:additional_results_bilayer}(a) shows the top and side views of a representative structure.

\subsubsection{Topological phase diagrams}

\begin{figure*}
    \centering
    \includegraphics[width = \linewidth]{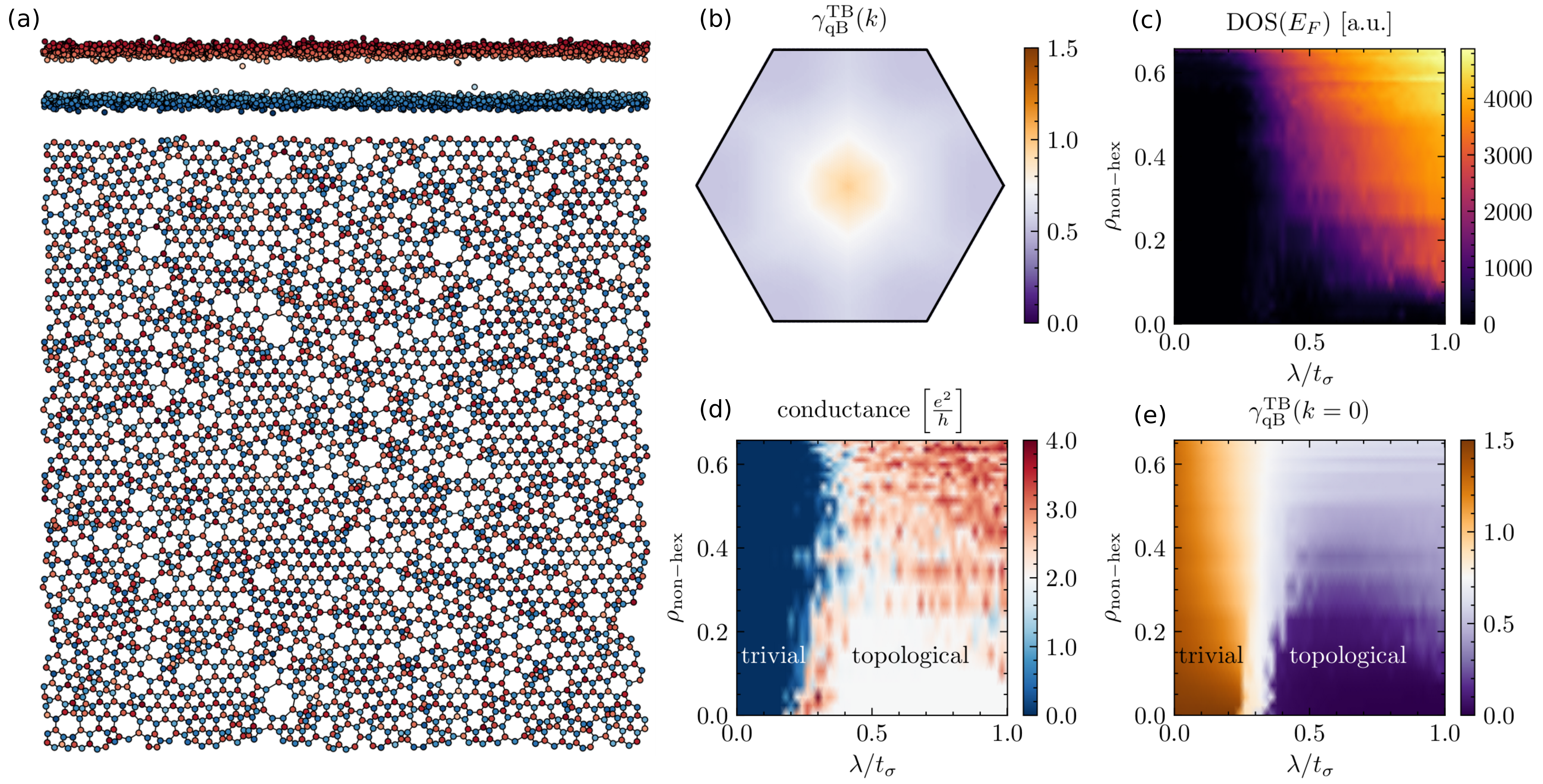}
    \caption{Bi bilayer tight-binding model structure and phase diagrams as a function of SOC $\lambda$ and amorphousness
    $\rho_{\text{non-hex}}$.
    (a) Top and side views of an example structure for amorphousness $\rho_{\text{non-hex}} = 0.53$ ($r=0.3$). Sites are colored according to their out-of-plane positions: red/blue indicates the effective sublattice, and the color intensity scales with the actual out-of-plane position. The positions in the out-of-plane direction have been rescaled by a factor 10 for visualization purposes. 
    (b) Momentum resolved tight-binding quasi-Bloch spillage for $\rho_{\text{non-hex}}=0.53$ ($r=0.3$) and SOC $\lambda = 0.7 t_{\sigma}$. These parameters are equal to those in Fig.~\ref{fig:DFTspillage}, with a change in color to match that of (e).
    (c) Phase diagram of the density of states at the Fermi level of the system with periodic boundary conditions.
    (d) Phase diagram of the two-terminal longitudinal conductance in the ``armchair'' ribbon configuration.
    (e) Phase diagram of the structural quasi-Bloch spillage $\gamma_{\mathrm{qB}}^{\mathrm{TB}}(\boldsymbol{k}=0)$ comparing the amorphous system with SOC $\lambda$ to a topological crystal with $\lambda=t_{\sigma}$.}
   \label{Sfig:additional_results_bilayer}
\end{figure*}

In this section, we study the topological phase diagram of the amorphous Bi bilayer tight-binding model \eqref{eq:H_Bi_bilayer}, and show that, as for Bimsuthene, the structural spillage correctly predicts the topological band inversion in this model.

Before analyzing the results, let us briefly review the current status regarding the topological characterization of crystalline Bi (111) bilayer. In the crystalline case with SOC, the Bi bilayer has been predicted to be a strong topological insulator \cite{murakami_quantum_2006,wada_localized_2011,liu_stable_2011,huang_nontrivial_2013}. Our model can also describe other materials with the same lattice, such as the antimony (111) bilayer. Due to the smaller SOC, the Sb bilayer becomes a strong topological insulator only when strained \cite{ares_recent_2018}. Therefore, our model in the crystalline case starts as a $\mathbb{Z}_2=0$ insulator for vanishing $\lambda$. A band inversion occurs at a finite value of $\lambda$, driving the system to a $\mathbb{Z}_2=1$ topological insulating phase. For the parameters used in this work (see Appendix \ref{app:bilayer_a}), this band inversion in the crystal occurs at $\Gamma$ for $\lambda \simeq 0.27t_{\sigma}$. 

As shown in Fig.~\ref{Sfig:additional_results_bilayer}(b), the structural quasi-Bloch spillage $\gamma_{\mathrm{qB}}^{\mathrm{TB}}(\boldsymbol{k})$ of the amorphous system with amorphousness $\rho_{\text{non-hex}}=0.53$ ($r=0.3$) and SOC $\lambda = 0.7 t_{\sigma}$ is maximum at $\boldsymbol{k}=0$, with a value $>0.75$, when the reference system is a trivial crystal with $\lambda = 0$. Per our topological criterion, explained in detailed in Appendix \ref{app:spillage_TB_criterion}, this indicates that there is still a band inversion at $\boldsymbol{k}=0$ in the presence of disorder.

Let us now analyze the topological phase diagram of the amorphous Bi bilayer tight-binding model. Figs.~\ref{Sfig:additional_results_bilayer}(d) and (e) show the conductance and the structural quasi-Bloch spillage, computed for a reference topological crystal with $\lambda = t_{\sigma}$, respectively, as a function of amorphousness, $\rho_{\text{non-hex}}$, and SOC, $\lambda$. Both phase diagrams show a transition from a trivial insulator at $\lambda \sim 0.2-0.3 t_{\sigma}$. 

First, note that the conductance shows a metallic region around the transition, also in the crystalline case. This is an artifact of the finite precision in computing the Fermi level with the kernel polynomial method, compounded with finite-size effects (see Appendix \ref{app:TB_details}). These effects also broaden the otherwise sharp transition in the structural spillage at low disorder. We have checked that this transition region is reduced upon increasing the kernel polynomial method precision and the system size. Note that these issues only appear as one approaches the transition, where the gap is increasingly small. For further related details, see also the discussion of Fig.~\ref{Sfig:qB_trP_qB_trPP} in Appendix \ref{app:spillage_TB}.

Let us now focus on the phases away from the transition. The trivial insulator phase at small $\lambda$, characterized by a vanishing conductance and a large spillage (since the reference crystal is topological), survives with amorphousness up to slightly higher $\lambda$ than in the crystalline case. On the other hand, the topological insulator phase, indicated by a quantized $2e^2/h$ conductance and a small spillage, only survives for small disorder, and the system seems to become slightly metallic for higher disorder. This metallic phase is further signaled by the finite density of states at the Fermi level shown in Fig.~\ref{Sfig:additional_results_bilayer}(c). Notice that, despite the absence of Rashba SOC in this model, the onsite $\lambda$ is already spin-non-conserving, and therefore a metallic phase can be the ground state. Nevertheless, we cannot discard the possibility that the metallic conductance is arising from finite-size effects with an Anderson localized bulk but with a localization length longer than the system sizes considered. A scaling study would be needed to discern the nature of this metallic conductance, but this lies beyond the scope of this work. In any case, the spillage is not specifically designed to capture such metallic feature, and it just indicates that the topological band inversion still (partially) occurs for high disorder. Nevertheless, the larger spillage at high disorder, where the disorder induces this potential metallic phase starting from a topological state, provides a signature for the partial loss of this band inversion. This partial melting of the band inversion is also compatible with the increasing density of states at the Fermi level shown in Fig.~\ref{Sfig:additional_results_bilayer}(c).

In summary, both conductance and spillage phase diagrams agree qualitatively and predict the topological phase transition. Quantitative differences only arise in the metallic regions, where the band inversion is just partial. As for bismuthene, we have also checked that the conductance with leads in the perpendicular direction and the spillages with other reference systems give similar results.

\subsubsection{\label{app:comparisonDFT}Comparison with DFT}

In this section, we comment on the comparison of the results of the previous section with the DFT results presented in the main text. In particular, let us compare the latter to the tight-binding results for the realistic SOC $\lambda \simeq 0.7 t_{\sigma}$. As shown in Fig.~\ref{fig:DFTspillage}, the structural spillage predicts a topological band inversion in the amorphous Bi bilayer in both DFT and tight binding. Both methods also agree on the fact that, above a certain disorder, the spectral gap closes (see Figs.~\ref{Sfig:additional_results_bilayer} and \ref{Sfig:pDOS_DFT}). Crucially, because we are forced to neglect the momentum scattering in the tight-binding approximation (see Appendix \ref{app:spillage_TB}), the structural spillage in DFT takes higher values and it is also less broad. Consequently, the structural spillage not only is a topological indicator compatible with DFT, but it works better in DFT than in tight-binding modeling.

\subsection{\label{app:TB_details}Calculation details}

This section describes in detail the methods used to solve the tight-binding models, and some related subtleties.

We use the Kwant software package \cite{groth_kwant_2014} to generate the tight-binding Hamiltonians and perform the calculations. To be able to treat larger system sizes, we apply the kernel polynomial method (KPM) \cite{weise_kernel_2006} to estimate the density of states (DOS) and the projector onto the occupied states. The projector is computed following the procedure of Ref.~\cite{Varjas20} and using plane waves as initial KPM vectors, which allows us to calculate the projector matrix elements $\langle \boldsymbol{p}\alpha |P | \boldsymbol{p} \beta \rangle$. We use a KPM energy resolution of $0.01 t_{\sigma}$ (645 moments) for the bismuthene structures, and of $0.005 t_{\sigma}$ (887 moments) for the bilayer ones. The DOS is computed by performing a KPM stochastic trace with 50 and 100 random vectors in the cases of bismuthene and bilayer, respectively. The system sizes considered are $21a \times 12\sqrt{3}a$ for the bismuthene case and $41a \times 24\sqrt{3}a$ for the Bi bilayer one. Both the resolution and the size of the Bi bilayer system are taken to be larger than those of bismuthene since the gap in the former case is smaller, and therefore finite-size effects are larger. Additionally, our model for the Bi bilayer displays some trivial edge states that affect the calculation of the Fermi level considerably.

\begin{figure}[!t]
    \centering
    \includegraphics[height=0.4\textwidth]{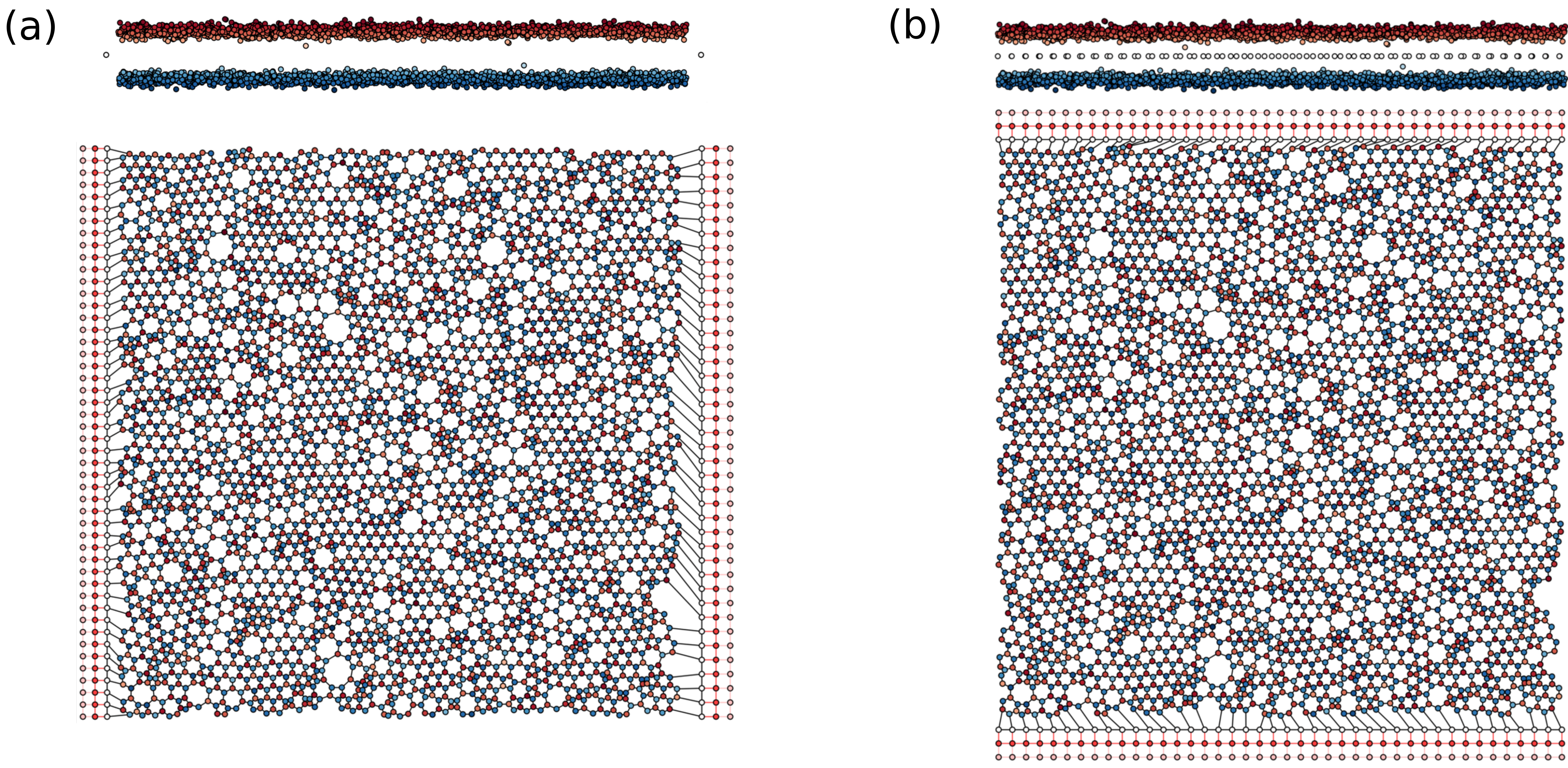}
    \caption{Examples of Bi bilayer systems with leads where conductance is calculated. (a) Top and side views of a system with leads in the $x$ axis, which would correspond to a zigzag ribbon in the crystalline case. (b) Top and side views of a system with leads in the $y$ axis, which would correspond to an armchair ribbon in the crystalline case.}
    \label{Sfig:leads}
\end{figure}

The structural quasi-Bloch spillage is computed in the systems with periodic boundary conditions using Eq.~\eqref{eq:qB-spillage-no-scatt-matrix-TB}, which reduces to Eq.~\eqref{eq:qB_spillage_TB_av_angles} in our models, since the crystalline phase has a honeycomb lattice. On the other hand, the conductance is determined with the Kwant software in the systems with open boundary conditions. In order to avoid possible artifacts arising from trivial edge states in some particular termination, the conductance is calculated using leads in both $x$ and $y$ directions, such that in the crystalline case the edges are zigzag and armchair, respectively. Since the aim of the conductance is to identify the insulating and topological insulating regions, which have a quantized conductance of 0 and $2e^2/h$, respectively, regardless of the shape of the leads, we use leads consisting of a 2D planar square lattice with nearest-neighbour hoppings such that their bandwidth is larger than that of the system. These leads are attached to all the atoms on the corresponding edge of the system. Fig.~\ref{Sfig:leads} shows two example configurations with the leads in the $y$ (armchair) and $x$ (zigzag) directions. 

Our Bi bilayer models, display at low disorder some trivial edge states close to the Fermi level over a wide range of values of SOC, which appear in both zigzag and armchair edges. These change the Fermi level of a finite system with open boundary conditions $E_F^{\text{open}}$ with respect to the one computed with periodic boundary conditions $E_F^{\text{periodic}}$. For the system sizes we are able to treat numerically the change in the Fermi level $E_F^{\text{open}}$ is enough for it to lie outside of the bulk gap, since the thermodynamic gap in the crystal is rather small ($\sim 0.1 t_{\sigma}$). Therefore, the conductance computed at $E_F^{\text{open}}$ in the crystal would show metallic regions even in the insulating and topological insulating phases due to this artifact. In order to avoid this issue, in the Bi bilayer systems we compute the conductance at $E_F^{\text{periodic}}$ determined with periodic boundary conditions. We note that this problem does not appear in the bismuthene models. It is also worth highlighting that the metallic phase observed at large SOC and disorder is not an artifact (see Appendix \ref{app:bilayer}), since we observe that the trivial edge states merge into bulk states in this region and therefore $E_F^{\text{periodic}} \simeq E_F^{\text{open}}$. 

Lastly, to compute the phase diagrams we only need a single disorder realization for each $r$. The reason is twofold. First, we noticed that for sufficiently large systems sizes, as the ones considered in this work, the fluctuations of the structural spillage for different disorder realizations are rather small. Indeed, they are smaller than the fluctuations in the conductance, which is another convenient feature for the use of the structural spillage in high-throughput searches for topological amorphous materials. Second, while extracting a precise topological phase diagram from the conductance would require a configuration average, it is not strictly necessary if we just aim to use it as a benchmark for the structural spillage.

\section{\label{app:DFT}DFT calculation details}

We performed Density Functional Theory (DFT) calculations using the projector augmented wave (PAW) formalism in the Vienna \textit{ab-initio} Simulation Package (VASP) \cite{VASP1, VASP2}. The exchange-correlation potentials were treated within the generalized gradient approximation (GGA) of Perdew-Burke-Ernzerbof (PBE) \cite{PBE1}. The wavefunctions were expanded in plane waves to an energy cutoff of 700 eV. SOC was added self-consistently for all calculations in which it was used. For supercell calculations, we performed Gamma point only calculations. For self-consistent calculations of the unit cell, we used a k-point grid of 21x21x1 with Gamma for the BZ sampling. We then sampled the 25 k-points ($\frac{n_1}{N_1}b_1+\frac{n_2}{N_2}b_2$) that would backfold to Gamma in the 5x5x1 supercell. To compare the same momenta between the unit cell and the supercell, the two must be commensurate and the supercell lattice vectors must be multiples of the unit cell lattice vectors. If this were not the case, one could linearly interpolate the coefficients of the supercell wavefunctions at the appropriate momenta from the closest supercell reciprocal lattice vectors.

Unlike in the tight-binding approximation, the structural spillage of Eq.~\eqref{eq:qB-spillage} can be directly implemented in DFT. Here, the overlap between two systems is well-defined irrespective of them having atoms at different positions. However, strictly speaking, the continuous set of plane waves is always overcomplete in any numerical scheme. Nevertheless, the structural spillage of Eq.~\eqref{eq:qB-spillage} is still well-defined in DFT implemented with both a plane-wave or a localized basis. On the one hand, plane-wave-based DFT codes feature discretized momenta (imposed by the periodic boundary conditions of the supercell) and a high-momentum cutoff. These features do not constitute any fundamental problem for comparing two systems with different atomic structures, as long as one has access to (or can interpolate) the information at the same momenta in both systems. On the other hand, implementations of DFT with a localized basis, such as Gaussian or hydrogenic orbitals, do not directly output the information in plane-wave momentum space. However, knowing the shape of the orbitals, a Fourier transform gives access to it, and no problem appears regardless of the atomic structure.

To calculate the structural spillage in DFT using Eq.~\eqref{eq:qB-spillage}, we extract the projector matrix elements on an orthonormal plane wave basis. The pseudo-wavefunctions generated with VASP are orthonormal with respect to an overlap operator \cite{PhysRevB.59.1758}.  Therefore, by using the PAW approach, we perform a transformation to an orthonormal basis that spans the same space as the full wavefunctions. Future improvements could use norm-conserving pseudopotentials, reconstructed full wavefunctions, or all-electron approaches. Besides imposing this orthonormality, we rearrange the wavefunction coefficient arrays of the amorphous supercell so that we compare the same momenta between both the amorphous supercell and the crystalline unit cell.

To corroborate that the spillage Eq.~\eqref{eq:qB-spillage} is correctly implemented, we compared a crystalline supercell to a crystalline unit cell, which should recover the exact Bloch spillage. In particular, we considered crystalline Bi$_2$Se$_3$ as well as crystalline BiTeI, and our method accurately diagnosed the band inversion in both systems. In crystalline Bi$_2$Se$_3$ a band inversion at Gamma leads to a topological insulator phase which results in a spillage value of 2.12 \cite{liu_spin-orbit_2014}. When comparing the crystalline Bi$_2$Se$_3$ supercell to the unit cell we obtain a spillage of 2.09 which exactly matches the result given by \textit{pymatgen} \cite{pymatgen}. For the case of disordered BiTeI, previous work showed that small amounts of disorder in the atomic positions cause the system to undergo a topological phase transitions from a trivial insulator (crystal) to a topological insulator (disordered) as a result of an induced band inversion \cite{corbae_structural_2021}. This is caused by the modified crystal field of the orbitals near the Fermi level which pushes these states closer together when disordered. In the latter case, all point group symmetries are broken but translational symmetry is still present.  In this case, we find a spillage value of 5.17 at the A point where the band inversion occurss, and values of 3.03 at other BZ points indicating there is a larger orbital spillage throughout the BZ. The method still captures the topological band inversion in this case and exactly matches the results given by pymatgen. 

\begin{figure}
    \centering
    \includegraphics[width=\textwidth]{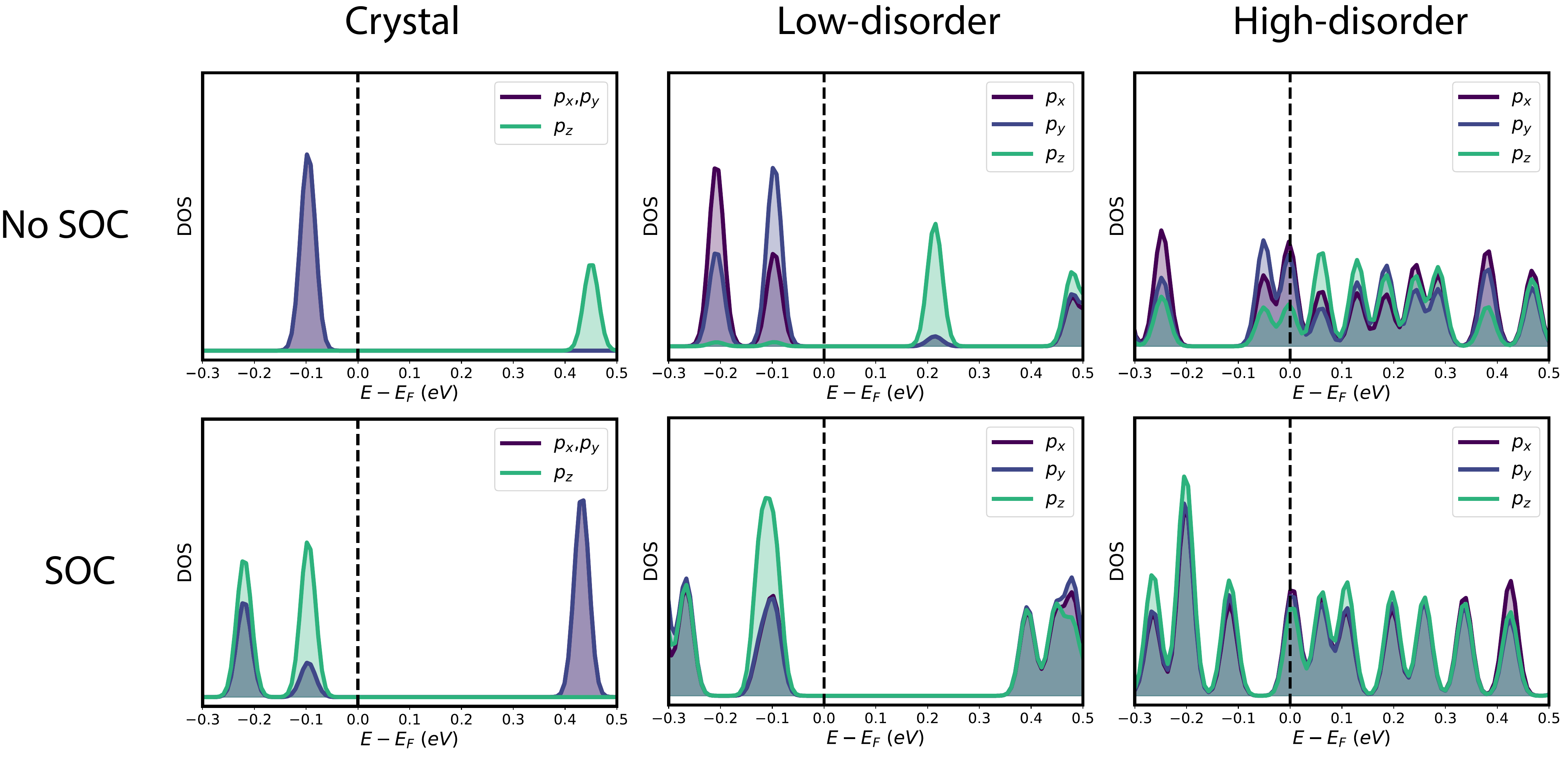}
    \caption{Orbital-resolved density of states (DOS) of the Bi (111) bilayer calculated with DFT, showing the contributions of the Bi $p$ orbitals near the Fermi level (indicated by a vertical dashed line). First row: DOS without SOC. Second row: DOS with SOC. Each column corresponds to a different structure: crystal in the the first column, low-disorder structure (standard deviation of $0.15 \angstrom$) in the second column, and high-disorder system (standard deviation of $0.30 \angstrom$) in the third column (see Fig.~\ref{fig:DFTstructure} in the main text for a real space view of these lattice structures). SOC drives a band inversion that occupies the $p_z$ orbital and empties the $p_{x,y}$ orbitals.}
    \label{Sfig:pDOS_DFT}
\end{figure}

\begin{figure}
    \centering
    \includegraphics[width=0.5\textwidth]{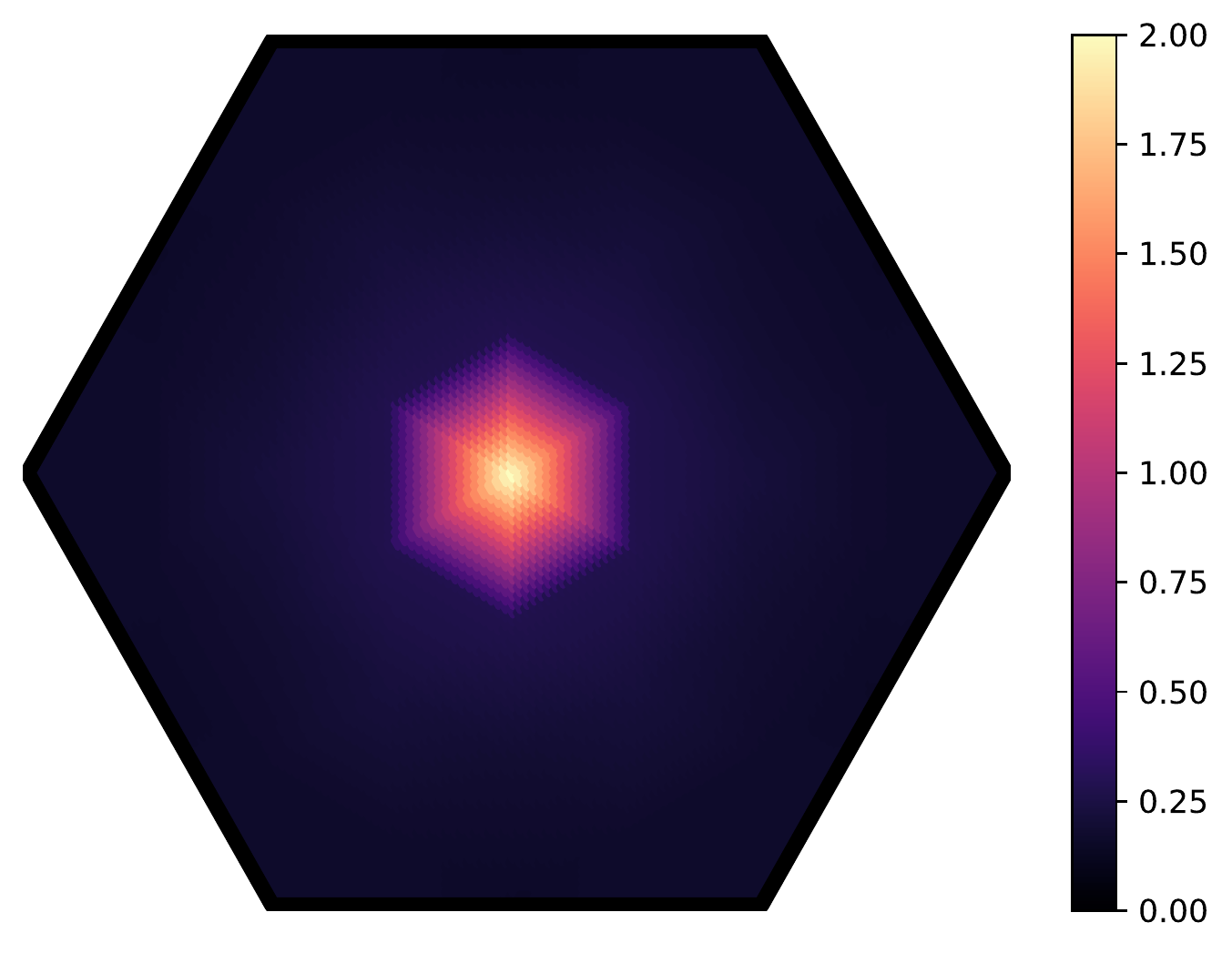}
    \caption{Calculated structural spillage of the crystalline Bi bilayer from DFT. The value of 2 at the Gamma point indicates that the crystalline Bi bilayer with SOC is topological.}
    \label{Sfig:xtal_spill}
\end{figure}

Finally, let us comment further on the results obtained for the Bi (111) bilayer. The disordered structures, shown in Fig.~\ref{fig:DFTstructure}, are obtained by randomly displacing the atoms from their high-symmetry crystal positions following a Gaussian distribution. We choose the standard deviations to be $0.15 \angstrom$ and $0.30 \angstrom$ for the low and high disorder systems, respectively. For standard deviations of $0.15 \angstrom$ the deviation from equilibrium position is small which preserves the bulk electronic gap while demonstrating our method works in the presence of disorder. Standard deviations of $0.30 \angstrom$ lead to an average atomic displacement of $0.41 \angstrom$ which is similar to atomic displacements seen in topological materials in the presence of disorder \cite{corbae_structural_2021}. The structural spillage, shown in Figs.~\ref{fig:DFTstructure} and \ref{Sfig:xtal_spill}, demonstrate that SOC drives a band inversion at the Gamma point with the result that all the  crystalline and the disordered structures are topologically non-trivial. This band inversion is confirmed by the density of states of Fig.~\ref{Sfig:pDOS_DFT}, which further illustrates that the band inversion occurs between the $p_z$ and the $p_{x,y}$ orbitals. Indeed, the crystal and the amorphous systems display an increased occupation of the $p_z$ orbital after SOC is included. Additionally, Fig.~\ref{Sfig:pDOS_DFT} illustrates that the Bi bilayer becomes metallic for sufficiently high structural disorder, in agreement to the tight-binding model (see section \ref{app:bilayer}). However, studying whether the amorphous system is extended or localized for strong disorder lies beyond the scope of this work.

\section{\label{app:spillage_TB}Defining the structural spillage in the tight-binding approximation}

\subsection{General remarks and motivation}

In the main text we use the tight-binding spillage as a benchmark, and argue that the structural spillage is most useful within DFT calculations. For completeness, in this appendix we give a pedagogical justification of Eq.~\eqref{eq:qB-spillage-no-scatt-matrix-TB} for computing the structural quasi-Bloch spillage in the tight-binding approximation. It is aimed to aid future studies in understanding the approximations that go into applying the structural spillage to tight-binding models, as alternative to topological markers. Thus it can be skipped by readers only interested in applying Eq.~\ref{eq:qB-spillage}.

Let us first highlight the problem of applying the general formulation of the structural spillage of Eq.~\eqref{eq:qB-spillage} in the tight-binding approximation. By tight-binding approximation we refer to the phenomenological tight-binding models where the only information about the wavefunctions is the position of their Wannier charge centers (and possibly their transformation properties under symmetries), but their spatial structure is unknown and therefore considered to be a Dirac delta. An implicit assumption of  Eq.~\eqref{eq:qB-spillage} is that the Hilbert space of the system is the whole real space (in addition to the spin space), in which the plane waves constitute an orthonormal basis. While this is applicable in DFT (see Appendix \ref{app:DFT}), it is not true in the tight-binding approximation, where the Hilbert space is just spanned by the positions of the Wannier charge centers (with the internal degrees of freedom of spin and orbital type). The fundamental problem for comparing two tight-binding systems with different lattice structures, as done by the structural spillage, stems from the fact that their Hilbert spaces are different, and therefore their overlap is ill-defined. When projected to the tight-binding Hilbert space, the plane waves constitute a non-orthogonal and overcomplete set. The overlap between these projected plane waves depends on the lattice structure, and therefore the usual formalism of non-orthogonal bases (see e.g. \cite{soriano_theory_2014}) cannot be applied. 

However, by using the plane waves and the approximations described in this Appendix, one can derive a physically motivated expression for the structural spillage in the tight-binding approximation, Eq.~\eqref{eq:qB-spillage-no-scatt-matrix-TB}. The line of the argument for solving this problem works as follows. The structural spillage \eqref{eq:qB-spillage} contains the matrix elements of the products of two projectors in the plane wave basis. By neglecting the momentum scattering, i.e., by assuming that these operators are diagonal in momentum space, the fundamental problem of the disorder-dependent plane-wave overlaps is circumvented. However, this introduces some new issues. To bypass these, we choose the solution which, in the crystalline limit, gives results closer to the exact ones. Our solution gives the exact results for the quantities containing matrix elements of just one projector. In the case of the structural spillage, which contains matrix elements of the product of two projectors, our results in the crystalline limit are not exact. However, we argue and numerically show for selected models that the results are similar in absolute value, and more importantly that the sharp changes in the spillage that signal topological transitions still show up.

In order to separately understand the different issues that appear in the tight-binding, let us first consider the simple case of a system whose corresponding crystalline limit has a single site per unit cell, where the majority of problems suffered by the structural spillage in the tight binding do not appear. Then, we will analyze the general multi-site case.

\subsection{System with a single site per unit cell}

\subsubsection{Setting the stage: crystalline system}

Consider a crystalline tight-binding system with $N_{\text{cells}}$ unit cells and one site per unit cell, i.e., only one Wyckoff position with multiplicity one is occupied by an atom, $N_{\mathrm{s/c}} = 1$. Therefore, the number of sites is the same as the number of cells, $N_{\text{sites}} = N_{\text{cells}}$. The number of internal degrees of freedom (orbitals and spins) at each site does not influence the discussion below, so we omit this internal index for simplicity in the notation. In the tight-binding approximation, Wannier functions  are unknown in real space, and therefore considered to be Dirac delta distributions, i.e., the Wannier function $|\phi_{\boldsymbol{R}}\rangle$ at the lattice site $\boldsymbol{R}$ has wavefunction:
\begin{equation}
    \phi_{\boldsymbol{R}}(\boldsymbol{r}) = \langle \boldsymbol{r} | \phi_{\boldsymbol{R}} \rangle = \delta(\boldsymbol{r}-\boldsymbol{R}).
\end{equation}
We will always assume that the Wannier functions are orthonormal:
\begin{equation}
    \langle \phi_{\boldsymbol{R}'} | \phi_{\boldsymbol{R}} \rangle = \delta_{\boldsymbol{R},\boldsymbol{R}'}.    
\end{equation}

The plane wave with momentum $\boldsymbol{p}$ projected to the tight-binding Hilbert space is a state with a phase $\boldsymbol{p} \cdot \boldsymbol{R}$ at the site $\boldsymbol{R}$, and normalized in the total volume of the system. Then, the Wannier functions in the plane wave basis read:
\begin{equation}
    \phi_{\boldsymbol{R}}(\boldsymbol{p}) = \langle \boldsymbol{p} | \phi_{\boldsymbol{R}} \rangle = \frac{1}{\sqrt{N_{\text{sites}}}} e^{-i\boldsymbol{p}\cdot\boldsymbol{R}}.
\end{equation}
Moreover, the Bloch states defined at crystal momentum $\boldsymbol{k}$ in the first BZ are:
\begin{equation}
    | \phi_{\boldsymbol{k}} \rangle = \frac{1}{\sqrt{N_{\text{cells}}}} \sum_{\boldsymbol{R}} e^{i\boldsymbol{k}\cdot\boldsymbol{R}} | \phi_{\boldsymbol{R}} \rangle,
\end{equation}
The overlap between the Bloch states and the plane waves is thus:
\begin{equation}
    \langle \boldsymbol{p} | \phi_{\boldsymbol{k}} \rangle = \frac{1}{N_{\text{sites}}} \sum_{\boldsymbol{R}} e^{i (\boldsymbol{k}-\boldsymbol{p})\cdot \boldsymbol{R}} =  \sum_{\boldsymbol{G}} \delta_{\boldsymbol{p},\boldsymbol{k}+\boldsymbol{G}},
    \label{eq:Bloch_state_in_pw_basis_single_atom}
\end{equation}
where $\boldsymbol{G}$ are the reciprocal lattice vectors, i.e., $\boldsymbol{G} \cdot \boldsymbol{R} / 2\pi \in \mathbb{Z}$. Therefore, all the BZs are exactly equivalent in a crystalline one-atom tight-binding, since 
\begin{equation}
    \langle \boldsymbol{k} + \boldsymbol{G} | \phi_{\boldsymbol{k}} \rangle = 1
    \label{eq:Bloch_state_in_pw_basis_single_atom_b}
\end{equation}
does not depend on $\boldsymbol{G}$. In other words, $\langle \boldsymbol{p} | \boldsymbol{p} + \boldsymbol{G} \rangle = 1$ for the crystal, i.e., both plane waves are projected to the same state, which is exactly the Bloch state at $\boldsymbol{k}$ too.

Finally, as a side remark, it is worth mentioning that even if there is a single site per unit cell, the BZs of a crystal are no longer equivalent if the orbitals have a finite spread in real space. Indeed, in this case, the overlap between the Bloch state and the plane waves is:
\begin{equation}
    \langle \boldsymbol{k} + \boldsymbol{G} | \phi_{\boldsymbol{k}} \rangle = \frac{1}{N_{\text{cells}}} \sum_{\boldsymbol{R}} e^{i\boldsymbol{k}\cdot\boldsymbol{R}} \langle \boldsymbol{k} + \boldsymbol{G} | \phi_{\boldsymbol{R}} \rangle = \frac{1}{N_{\text{cells}}} \sum_{\boldsymbol{R}} e^{-i\boldsymbol{G}\cdot\boldsymbol{R}} \langle \boldsymbol{k} + \boldsymbol{G} | \phi_{0} \rangle = \phi_0(\boldsymbol{k}+\boldsymbol{G}),
\end{equation}
where $\phi_0(\boldsymbol{k}+\boldsymbol{G})$ is the Fourier transform of the orbital located at the origin, which is generically not constant.

\subsubsection{Spillage comparing two crystals}

Let us remember that plane waves are an overcomplete set in the tight-binding Hilbert space. In this single-site case, the Hilbert space dimension is $N_{\text{sites}}$, which is the number of linearly independent plane waves needed for a basis. One possible choice is selecting all the $N_{\text{cells}} = N_{\text{sites}}$ momenta in one BZ (e.g. the first BZ). These are linearly independent and orthogonal in the crystalline case (and also for an amorphous structure in the infinite size limit). Therefore, this choice constitutes an orthonormal basis. Therefore, in this basis we can directly apply Eq.~\eqref{eq:qB-spillage-crystal} for the spillage, choosing to compare two crystals, with the particularity that the sums over reciprocal lattice vectors $\boldsymbol{G}$ disappear since there is only one in the basis. The key difference from the general multi-site case is that observables are the same irrespective of the BZ where the momenta for the basis are chosen, i.e., irrespective of the $\boldsymbol{G}$ chosen in the basis. Moreover, thanks to the equivalence between plane waves and Bloch states in this single-site case, observables projected to a plane wave $\boldsymbol{p}$ are equal to the crystalline quantities computed at Bloch momentum $\boldsymbol{k} = \boldsymbol{p} \text{ }\mathrm{mod}\text{ } \boldsymbol{G}$. In particular, the quasi-Bloch spillage \eqref{eq:qB-spillage}, which is equal to the Bloch spillage because we are comparing two crystals, is also equal to the quasi-Bloch spillage without scattering \eqref{eq:qB-spillage-no-scatt-matrix-TB} in this crystalline one-site case.

\subsubsection{Structural spillage comparing an amorphous system to a crystal\label{app:spillage_TB_single_site_subsec}}

The previous basis choice is also orthonormal for an amorphous system in the infinite-size limit. Consequently, unlike in the multi-site case that will be analyzed in the next section, the issue of the overlap between plane waves being different for the amorphous and crystalline systems does not appear. Therefore, the structural quasi-Bloch spillage including scattering of Eq.~\eqref{eq:qB-spillage} can also be applied for comparing the amorphous structure with a crystalline one in this single-site tight-binding case (again the sums over reciprocal lattice vectors $\boldsymbol{G}$ drop out in this single-site case). As mentioned in the previous section, when comparing two crystals with a single site per unit cell, the quasi-Bloch spillage including scattering of Eq.~\eqref{eq:qB-spillage} coincides with the one without scattering of Eq.~\eqref{eq:qB-spillage-no-scatt-matrix-TB}. This is no longer true when comparing an amorphous structure to a crystal, since the scattering resummation over $\boldsymbol{k}'$ in the amorphous projector, which is carried out in Eq.~\eqref{eq:qB-spillage}, is neglected in Eq.~\eqref{eq:qB-spillage-no-scatt-matrix-TB}.

Now, although the structural quasi-Bloch spillage including scattering of Eq.~\eqref{eq:qB-spillage} could in principle be applied, this would entail a high computational cost. Indeed, other methods to indicate the topology in the tight-binding would be equally efficient (such as the local topological markers \cite{kitaev_anyons_2006,prodan_non-commutative_2010,bianco_mapping_2011,hannukainen_local_2022}), questioning the usefulness of the structural spillage applied to a tight-binding model. Therefore, to implement efficiently the structural spillage, we assume the no-scattering approximation of Eq.~\eqref{eq:qB-spillage-no-scatt-matrix-TB}. Because we neglect the scattering resummation over $\boldsymbol{k}'$, the structural spillage of Eq.~\eqref{eq:qB-spillage-no-scatt-matrix-TB} becomes much more computationally efficient.

However, an important inconvenience arising from neglecting the scattering is that the spillage depends on the BZ where the momenta for the plane wave basis are chosen. This is because momenta from different crystalline BZs will no longer lead to equivalent results in the amorphous system, unlike in the single-site crystal. In fact, $| \boldsymbol{p} + \boldsymbol{G} \rangle$ and $| \boldsymbol{p} \rangle$ no longer project to the same state ($\langle \boldsymbol{p} | \boldsymbol{p} + \boldsymbol{G} \rangle = 0$ for the amorphous case in the infinite size limit), and the quantities projected in $| \boldsymbol{p} + \boldsymbol{G} \rangle$ differ from those projected onto $| \boldsymbol{p} \rangle$. 

This problem raises the question of how to compute correctly the structural spillage in the no-scattering approximation between an amorphous material and a crystal, even in this single-site case. Although there is no unique answer, we now provide a justification for using momenta just in the first BZ. The tight binding has no information about the spatial extent of the orbitals, although we know that they are exponentially localized around the atom. Therefore, the tight-binding approximation captures well long-distance physics, but there is a short-distance-cutoff below which the tight-binding results are no longer reliable. It is reasonable to assume that this cutoff is of the order of the nearest-neighbour distance $r_{\mathrm{nn}}$, which coincides with the lattice constant $a$ in the crystalline single-site tight-binding. Therefore, only plane-wave momenta below $\sim 2\pi/a$ are reliable. Consequently, the quasi-Bloch spillage computed just with plane-wave momenta in the first BZ is a sensible option (optionally, one could average over the first BZ and second BZs). Considering just the first BZ, the structural quasi-Bloch spillage without scattering reads 
\begin{equation}
    \gamma_{\mathrm{qB}}^{\text{single-site-TB}}(\boldsymbol{k}) = \frac{1}{2} \text{tr} \left[ \left( P_{\boldsymbol{k}} - \tilde{P}_{\boldsymbol{k}} \right)^2 \right],
\label{eq:qB-spillage-no-scatt-matrix-TB-single-site}
\end{equation}
which is just Eq.~\eqref{eq:qB-spillage-no-scatt-matrix-TB} in the single-site case because, as mentioned before, all BZs are equivalent in the crystal, and therefore there is a single type of BZ, $N_{\mathrm{BZs}} = 1$.

\subsection{System with several sites per unit cell}

In this section, we will show that if there are more than one site in the unit cell, then a phase factor depending on the relative positions of the sites appears in the observables. Unlike in the single-site case, this leads to some BZs being inequivalent in the crystal, requiring us to upgrade the single-site structural spillage Eq.~\eqref{eq:qB-spillage-no-scatt-matrix-TB-single-site}.

\subsubsection{Crystal: definitions and types of Brillouin zones \label{app:spillage_TB_multi_crystal_types_BZ}}

Consider a crystal with $N_{\text{cells}}$ unit cells at positions $\boldsymbol{R}$ and $N_{\text{s/c}}$ sites per unit cell at positions $\boldsymbol{t}_A$ with respect to the center of the cell $\boldsymbol{R}$, so that the total number of sites is $N_{\text{sites}} = N_{\text{cells}} \cdot N_{\text{s/c}}$. The Bloch states with a definite sublattice are, therefore:
\begin{equation}
    | \phi_{\boldsymbol{k}}^{A} \rangle = \frac{1}{\sqrt{N_{\text{cells}}}} \sum_{\boldsymbol{R}} e^{i\boldsymbol{k}\cdot(\boldsymbol{R}+\boldsymbol{t}_A)} | \phi_{\boldsymbol{R}}^{A} \rangle.
\end{equation}
The projection of the Wannier functions onto plane-waves reads:
\begin{equation}
    \phi_{\boldsymbol{R}}^{A}(\boldsymbol{p}) = \langle \boldsymbol{p} | \phi_{\boldsymbol{R}}^{A} \rangle = \frac{1}{\sqrt{N_{\text{sites}}}} e^{-i\boldsymbol{p}\cdot(\boldsymbol{R}+\boldsymbol{t}_A)}.
\end{equation}
Therefore, the overlap between the Bloch states and the plane waves is:
\begin{equation}
    \langle \boldsymbol{k} + \boldsymbol{G} | \phi_{\boldsymbol{k}}^{A} \rangle = \frac{1}{\sqrt{N_{\text{s/c}}}} e^{-i\boldsymbol{G}\cdot\boldsymbol{t}_A}.
\end{equation}
However, the band eigenvectors are combinations of these Bloch states in different sublattices:
\begin{equation}
    |\psi_{\boldsymbol{k}}^n \rangle = \sum_{A} c_{\boldsymbol{k}}^{nA} | \phi_{\boldsymbol{k}}^{A} \rangle,
\end{equation}
and, therefore, their overlap with the plane waves reads:
\begin{equation}
    \langle \boldsymbol{k} + \boldsymbol{G}|\psi_{\boldsymbol{k}}^n \rangle = \frac{1}{\sqrt{N_{\text{s/c}}}} \sum_{A} c_{\boldsymbol{k}}^{nA} e^{-i\boldsymbol{G}\cdot\boldsymbol{t}_A},
\end{equation}

Let us now show that observables projected to a plane wave with momentum $\boldsymbol{p} = \boldsymbol{k} + \boldsymbol{G}$ depend on the phase factors $e^{-i\boldsymbol{G}\cdot\boldsymbol{t}_{AB}}$, where $\boldsymbol{t}_{AB} = \boldsymbol{t}_{A} - \boldsymbol{t}_{B}$ are the relative positions of the different sublattices. For concreteness, let us start considering the simplest observable, that will be a building block for e.g. the spillage: the projector onto band $n$ at crystal momentum $\boldsymbol{k}$, $P^n_{\boldsymbol{k}} = \big| \psi_{\boldsymbol{k}}^n \rangle \langle \psi_{\boldsymbol{k}}^n \big|$:
\begin{equation}
    \langle \boldsymbol{k} + \boldsymbol{G} \big| P^n_{\boldsymbol{k}} \big| \boldsymbol{k} + \boldsymbol{G} \rangle = 
    \big| \langle \boldsymbol{k} + \boldsymbol{G} \big| \psi_{\boldsymbol{k}}^n \rangle \big|^2 = 
    \frac{1}{N_{\text{s/c}}} \sum_{A, B} c_{\boldsymbol{k}}^{nA} \left(c_{\boldsymbol{tk}}^{nB}\right)^* e^{-i\boldsymbol{G}\cdot\boldsymbol{t}_{AB}} =
    \frac{1}{N_{\text{s/c}}} \left[ 1 + \sum_{A \neq B} c_{\boldsymbol{k}}^{nA} \left(c_{\boldsymbol{k}}^{nB}\right)^* e^{-i\boldsymbol{G}\cdot\boldsymbol{t}_{AB}} \right],
\label{eq:Pnk_projected_kG}
\end{equation}
which is different from $\mathrm{tr}\left[ P^n_{\boldsymbol{k}} \right] = 1$ in general. These phase factors, which depend on $\boldsymbol{G}$, lead to at least some BZs being inequivalent even if the orbitals are still Dirac deltas. Therefore, the types of BZs in the multi-site crystal can be classified by the set of phase factors $\left\{e^{-i\boldsymbol{G}\cdot\boldsymbol{t}_{AB}}\right\}$. In general, some BZs become inequivalent whenever there is structure inside the unit cell, irrespective of whether it comes from spatially-extended orbitals or from several sites. 

\begin{figure}
    \centering
    \includegraphics[width=0.3\textwidth]{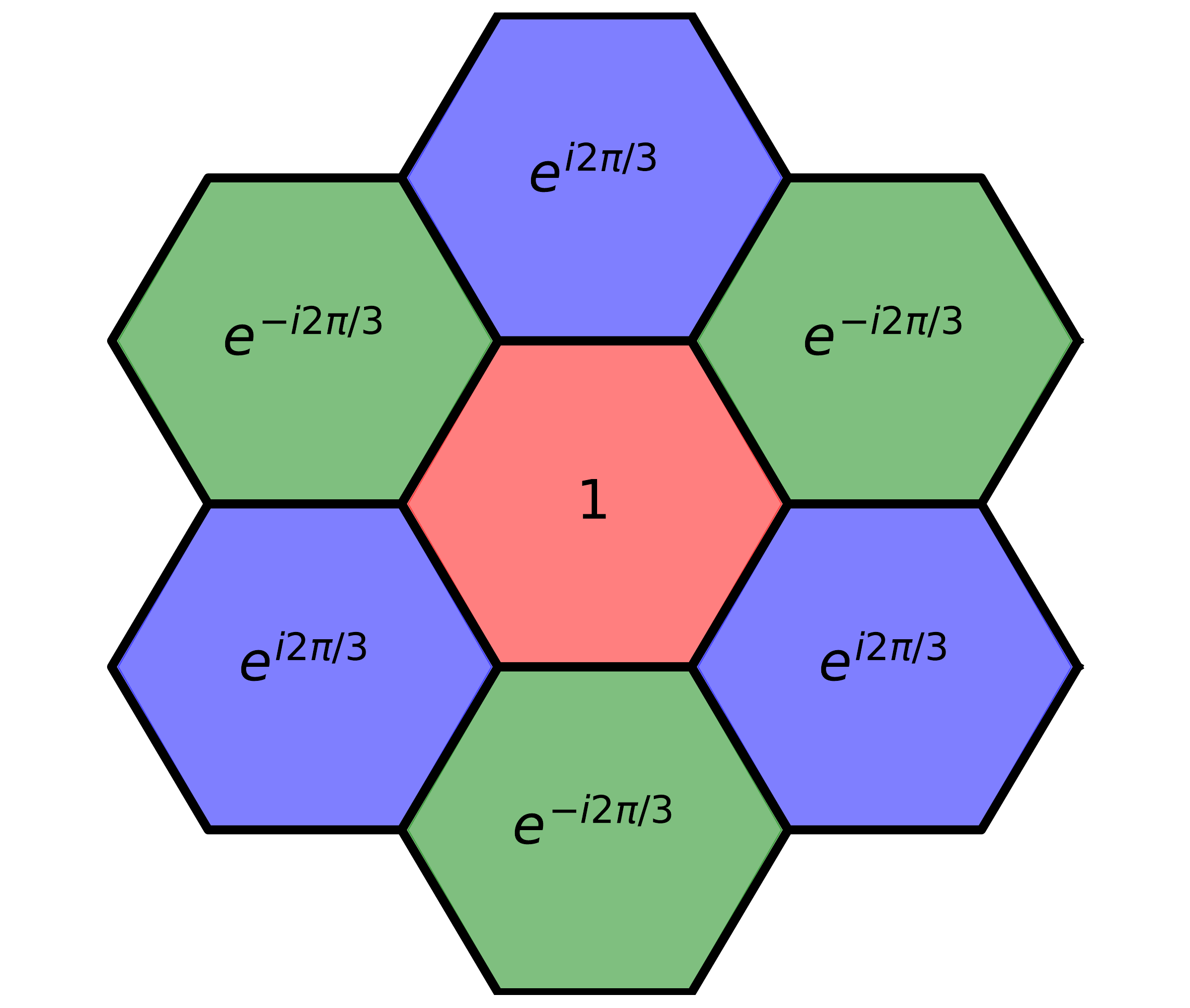}
    \caption{BZ types for the honeycomb lattice. Colors are different for each type. Red corresponds to $a=0$ mod 3, and therefore a phase $e^{-i\boldsymbol{G} \cdot \boldsymbol{t}_{AB}} = e^{i a 2\pi/3} = 1$. Blue represents $a=1$ mod 3, i.e., a phase $e^{i 2\pi/3}$. Finally, green refers to $a=2$ mod 3, i.e., a phase $e^{-i 2\pi/3}$.}
    \label{Sfig:BZ_types_honeycomb}
\end{figure}

As an example, consider the honeycomb lattice, where there are $N_{\mathrm{s/c}} = 2$ sublattices $A$ and $B$ such that $\boldsymbol{t}_{AB} = - a \left[0, 1/\sqrt{3}\right]$. The reciprocal lattice basis vectors are $\boldsymbol{G}_1 = 4 \pi / \sqrt{3} a \left[ \sqrt{3}/2 , 1/2 \right]$, and $\boldsymbol{G}_2 = 4 \pi / \sqrt{3} a \left[ 0 , 1 \right]$. A general reciprocal lattice vector $\boldsymbol{G} = n_1 \boldsymbol{G}_1 + n_2 \boldsymbol{G}_2$, with $n_1,n_2 \in \mathbb{Z}$, satisfies $\boldsymbol{G} \cdot \boldsymbol{t}_{AB} = - 4\pi/3 (2n_2+n_1) = 2\pi/3 \cdot 2 (2n_2+n_1)$. Therefore, $e^{-i\boldsymbol{G} \cdot \boldsymbol{t}_{AB}} = e^{i a 2\pi/3}$, with $a \in \mathbb{Z}_3$, so there are $N_{\mathrm{BZs}} = 3$ different types of BZs depending on the value of this phase factor. If we consider all possible momenta, from zero to infinity, then the multiplicity in momentum space of each type of BZ is the same. On the other hand, if we only consider momenta up to a cutoff $p_{\mathrm{max}}$, then the multiplicity in momentum space of each type of BZ can be different. Fig.~\ref{Sfig:BZ_types_honeycomb} shows the type of the first BZ and the six nearest-neighbour second BZs. Note that the first BZ has $\boldsymbol{G}=0$, and therefore it is always characterized by $a=0$, i.e., by a phase $e^{-i\boldsymbol{G} \cdot \boldsymbol{t}_{AB}} = e^{i a 2\pi/3} = 1$.

\subsubsection{Crystal: recovering the exact results using plane waves \label{app:spillage_TB_multi_crystal_exact}}

We now ask the question of how to recover the exact values of the observables in the crystalline tight binding, this time using the plane waves. We also keep in mind that we want to later extend our definitions to the amorphous case. 

First, we have to choose a basis of plane waves for this crystalline multi-site case. The tight-binding Hilbert space has dimension $N_{\mathrm{sites}} = N_{\mathrm{s/c}} \cdot N_{\mathrm{cells}}$. Therefore, a possibility is to select $N_{\mathrm{cells}}$ plane waves in $N_{\mathrm{s/c}}$ inequivalent BZs. Decomposing the plane-wave momenta as $\boldsymbol{p} = \boldsymbol{k} + \boldsymbol{G}$, we find that plane waves with different $\boldsymbol{k}$ are orthogonal. However, in contrast to the single-site case, plane waves with the same $\boldsymbol{k}$ but differing in a reciprocal lattice vector $\boldsymbol{G}$ are generically neither orthogonal nor equivalent in the crystalline case. 
It is only when the differing reciprocal lattice vector $\boldsymbol{G}$ verify $\left\{e^{-i\boldsymbol{G}\cdot\boldsymbol{t}_{AB}}\right\} = \left\{1\right\}$, i.e., when the BZs are equivalent, that the projected plane waves are equivalent states.

For instance, in the honeycomb lattice, where $N_{\mathrm{s/c}} = 2$, we can choose the basis in the first BZ ($\boldsymbol{G}_0 = 0$) and in the $\boldsymbol{G}_1 = 4\pi/\sqrt{3}a(0,1)$ BZ. In this example, the overlap between plane waves is $|\langle \boldsymbol{k} + \boldsymbol{G}_0 | \boldsymbol{k} + \boldsymbol{G}_1 \rangle| = |\langle \boldsymbol{k}| \boldsymbol{k} + \boldsymbol{G}_1 \rangle| = 0.5$. Therefore, we have to use the formalism of non-orthogonal bases (see, e.g., Ref.~\cite{soriano_theory_2014}) and properly modify the quasi-Bloch spillage of Eq.~\eqref{eq:qB-spillage-general}. Within this formalism, the closure relation reads:
\begin{equation}
    \mathbbm{1} = \sum_{\boldsymbol{k}} \sum_{\boldsymbol{G}\boldsymbol{G}'} \big| \boldsymbol{k} + \boldsymbol{G} \rangle \left(S^{-1}\right)_{\boldsymbol{G},\boldsymbol{G}'} \langle \boldsymbol{k} + \boldsymbol{G}' \big|,
\end{equation}
where the overlap matrix is defined as $S_{\boldsymbol{G},\boldsymbol{G}'} = \langle \boldsymbol{k} + \boldsymbol{G} \big| \boldsymbol{k} + \boldsymbol{G}' \rangle$, which depends only on the difference $\boldsymbol{G}'-\boldsymbol{G}$. Also, the sums over the reciprocal lattice vectors $\boldsymbol{G}$ run over the $N_{\mathrm{s/c}}$ BZs chosen in the basis. In the previous example of the honeycomb lattice, they would run over $\boldsymbol{G}_0 = 0$ and $\boldsymbol{G}_1 = 4\pi/\sqrt{3}a(0,1)$. Using this expression for the closure relation, we can derive the expressions for the observables in this non-orthogonal plane-wave basis. For example, the trace of the projector onto band $n$ at crystal momentum $\boldsymbol{k}$, $\mathrm{tr}\left[ P^n_{\boldsymbol{k}} \right]$, becomes
\begin{equation}
    \mathrm{tr}\left[ P^n_{\boldsymbol{k}} \right]_{\text{non-orth}} = \sum_{\boldsymbol{G}\boldsymbol{G}'} \langle \boldsymbol{k} + \boldsymbol{G} \big| P^n_{\boldsymbol{k}} \big| \boldsymbol{k} + \boldsymbol{G}' \rangle \left(S^{-1}\right)_{\boldsymbol{G}',\boldsymbol{G}},
\label{eq:trPnk_non-orthogonal}
\end{equation}
Importantly, Eq.~\eqref{eq:trPnk_non-orthogonal} recovers the expected crystalline value $\mathrm{tr}\left[ P^n_{\boldsymbol{k}} \right] = 1$, irrespective of the chosen plane-wave basis. Furthermore, in this non-orthogonal basis, the quasi-Bloch spillage is given by the appropriate generalization of Eq.~\eqref{eq:qB-spillage-general}: 
\begin{equation}
\begin{split}
    \gamma_{\mathrm{qB}}^{\text{non-orth}}(\boldsymbol{k}) = \frac{1}{2} \sum_{\boldsymbol{k}'} \sum_{\boldsymbol{G}_1\boldsymbol{G}_2\boldsymbol{G}_3\boldsymbol{G}_4} \sum_{\alpha\beta} & \left[ 
	P^{\alpha\beta}_{\boldsymbol{k}+\boldsymbol{G}_1,\boldsymbol{k}'+\boldsymbol{G}_2} 
	\left(S^{-1}\right)_{\boldsymbol{G}_2,\boldsymbol{G}_3}
	P^{\beta\alpha}_{\boldsymbol{k}'+\boldsymbol{G}_3,\boldsymbol{k}+\boldsymbol{G}_4}
	\left(S^{-1}\right)_{\boldsymbol{G}_4,\boldsymbol{G}_1} - \right. \\
	& \left. - P^{\alpha\beta}_{\boldsymbol{k}+\boldsymbol{G}_1,\boldsymbol{k}'+\boldsymbol{G}_2} 
	\left(S^{-1}\right)_{\boldsymbol{G}_2,\boldsymbol{G}_3}
	\tilde{P}^{\beta\alpha}_{\boldsymbol{k}'+\boldsymbol{G}_3,\boldsymbol{k}+\boldsymbol{G}_4}
	\left(S^{-1}\right)_{\boldsymbol{G}_4,\boldsymbol{G}_1}
	\right]	+ \left[P \leftrightarrow \tilde{P} \right].
\end{split}
\label{eq:qB-spillage-non-orthogonal} 
\end{equation}
Crucially, when comparing two crystals, Eq.~\eqref{eq:qB-spillage-non-orthogonal} exactly recovers the Bloch spillage, regardless of the plane wave basis chosen.

\subsubsection{Comparing an amorphous system to a crystal using the structural spillage: no-scattering approximation\label{app:spillage_TB_multi_no_scatt}}

Let us now try to compute the structural spillage between a crystalline and an amorphous structure. Aside from the issues already discussed for the single-site case, here is where comparing two tight bindings with sites at different positions becomes problematic. The reason is that overlap between the plane waves is different in the crystal and in the amorphous cases. In the crystal, as discussed in section \ref{app:spillage_TB_multi_crystal_types_BZ}, some plane waves $\big| \boldsymbol{p} + \boldsymbol{G} \rangle$ are different states from $\big| \boldsymbol{p} \rangle$, yet their overlap is non-zero, $\langle \boldsymbol{p} \big| \boldsymbol{p} + \boldsymbol{G} \rangle \neq 0$. In the amorphous system, in the limit of infinite size, all plane waves are inequivalent (as in the single-site case), and more significantly, they are orthogonal. In the structural spillage of Eq.~\eqref{eq:qB-spillage-non-orthogonal}, the crystalline and the amorphous projector appear sandwiched between the overlap matrices, but this overlap depends on the system. Therefore, we cannot apply the previous non-orthogonal formalism. 

As explained in the main text, this issue can be avoided by neglecting the momentum scattering, i.e., by setting $\boldsymbol{k}'=\boldsymbol{k}$ and $\boldsymbol{G}'=\boldsymbol{G}$ in Eq.~\eqref{eq:qB-spillage-general}. Such approximation has been used previously to determine the topology of an amorphous system using other methods such as the effective Hamiltonian approach \cite{Varjas2019,Marsal2020}. It is also inspired by the fact that continuous translational symmetry is recovered after averaging over different disorder realizations. 

Let us now write the expressions for the projector and the spillage within this approximation. On the one hand, the trace of the projector into band $n$ at crystal momentum $\boldsymbol{k}$ simplifies to:
\begin{align}
    \mathrm{tr}\left[ P^n_{\boldsymbol{k}} \right]_{\text{no scatt}} = \sum_{\boldsymbol{G}} \langle \boldsymbol{k} + \boldsymbol{G} \big| P^n_{\boldsymbol{k}} \big| \boldsymbol{k} + \boldsymbol{G} \rangle,
\label{eq:trPnk_no-scatt}
\end{align}
where the sums over the reciprocal lattice vectors $\boldsymbol{G}$ again run over the $N_{\mathrm{s/c}}$ BZs chosen in the plane wave basis. On the other hand, the corresponding expression for the structural quasi-Bloch spillage without scattering, which is obtained by setting $\boldsymbol{k}'=\boldsymbol{k}$ and $\boldsymbol{G}'=\boldsymbol{G}$ in Eq.~\eqref{eq:qB-spillage-general}, reads:
\begin{equation}
    \gamma_{\mathrm{qB}}^{\text{no scatt}}(\boldsymbol{k}) = \frac{1}{2} \sum_{\boldsymbol{G}} \text{tr} \left[ \left( P_{\boldsymbol{k}+\boldsymbol{G}} - \tilde{P}_{\boldsymbol{k}+\boldsymbol{G}} \right)^2 \right],
\label{eq:qB-spillage-no-scatt-matrix}
\end{equation}
where the trace acts over the internal degrees of freedom $\alpha$, and, as in the main text, $P_{\boldsymbol{p}}^{\alpha\beta} = \langle \boldsymbol{p}| P | \boldsymbol{p} \rangle$.  Eq.~\eqref{eq:qB-spillage-no-scatt-matrix} is not yet the definite expression of Eq.~\eqref{eq:qB-spillage-no-scatt-matrix-TB} for the structural spillage in the tight-binding approximation, since it still suffers from a problem that we detail below.

\subsubsection{Taking into account different types of Brillouin zones}

In contrast to the single-site case, the values of the observables computed within this no-scattering approximation depend on the BZs chosen in the basis even in the crystal. The reason is the presence of different types of BZs (see Appendix \ref{app:spillage_TB_multi_crystal_types_BZ}). In this section, we will provide a method to circumvent this issue based on the condition that, when applied to crystals, it leads to values as close as possible to the exact crystalline values, where rigorous proofs exist~\cite{liu_spin-orbit_2014}.

In short, our solution consists of computing a observable without scattering, performing an average over the $N_{\mathrm{BZs}}$ different types of BZs, and then multiplying by the number of sites per unit cell $N_{\mathrm{s/c}}$ in the crystal. First, let us show that our proposal recovers the correct crystalline result for the observables that depend only on one projector. Indeed, the BZ-averaged Eq.~\eqref{eq:trPnk_no-scatt} representing the trace of the projector into the band $n$ at crystal momentum $\boldsymbol{k}$ becomes:
\begin{equation}
    \mathrm{tr}\left[ P^n_{\boldsymbol{k}} \right]_{\text{no scatt}}^{\text{BZ av}} = \frac{N_{\text{s/c}}}{N_{\mathrm{BZs}}} \sum_{a \in \mathrm{BZs}} \langle \boldsymbol{k} + \boldsymbol{G}_a \big| P^n_{\boldsymbol{k}} \big| \boldsymbol{k} + \boldsymbol{G}_a \rangle = 1 + \sum_{A \neq B} c_{\boldsymbol{k}}^{nA} \left(c_{\boldsymbol{k}}^{nB}\right)^* \left[ \frac{1}{N_{\mathrm{BZs}}} \sum_{a \in \mathrm{BZs}} e^{-i\boldsymbol{G}_a\cdot\boldsymbol{t}_{AB}} \right] = 1,
\label{eq:trP_avBZ}
\end{equation}
where the sum over $a$ runs over a representative BZ of each type, and we have used Eq.~\eqref{eq:Pnk_projected_kG} and the fact that the term inside the square brackets vanishes identically for $A\neq B$. If there is a finite number $N_{\mathrm{BZs}}$ of BZ types, this term vanishes because the $N_{\mathrm{BZs}}$ phases $e^{-i\boldsymbol{G}_a\cdot\boldsymbol{t}_{AB}}$ are the $1/N_{\mathrm{BZs}}$ roots of unity. If there are infinite BZ types, which might occur, e.g., if the sites are located at a generic nonsymmetric Wyckoff position incommensurate with the reciprocal lattice vectors, then this term vanishes due to the infinite sum of a continuum of phases. In the example of the honeycomb lattice, where $N_{\mathrm{BZs}}=3$ and $e^{-i\boldsymbol{G}_a \cdot \boldsymbol{t}_{AB}} = e^{i a 2\pi/3}$ with $a \in \mathbb{Z}_3$ if $A\neq B$, and $e^{-i\boldsymbol{G}_a \cdot \boldsymbol{t}_{AB}} = 1$ if $A=B$, we obtain, as expected:
\begin{equation}
    \frac{1}{3} \sum_{a=0,1,2} e^{-i\boldsymbol{G}_a\cdot\boldsymbol{t}_{AB}} = \delta_{AB}.
\end{equation}

\begin{figure}[!t]
    \centering
    \includegraphics[height=0.3\textwidth]{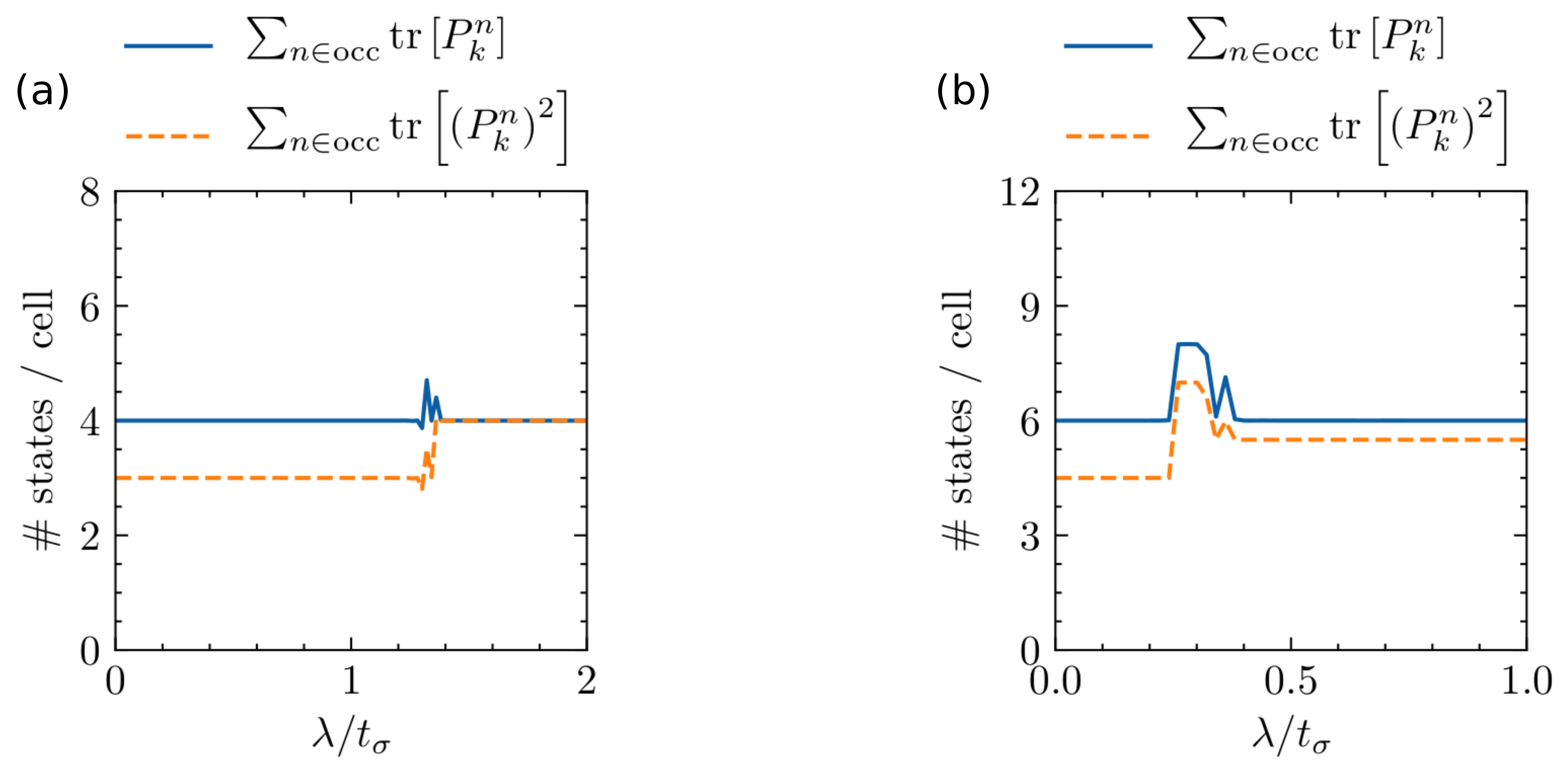}
    \caption{Sum over occupied bands of the trace of one and two projectors, $\sum_{n\in \mathrm{occ}} \mathrm{tr}\left[ P^n_{\boldsymbol{k}} \right]_{\text{no scatt}}^{\text{BZ av}}$ and $\sum_{n\in \mathrm{occ}} \mathrm{tr}\left[ \left( P^n_{\boldsymbol{k}} \right)^2 \right]_{\text{no scatt}}^{\text{BZ av}}$, as a function of onsite SOC, computed using the formalism of Eqs.~\eqref{eq:trP_avBZ} and \eqref{eq:trPP_avBZ} at $\boldsymbol{k}=0$. (a) Bismuthene crystal. (b) Bi bilayer crystal. On the one hand, the filling $\sum_{n\in \mathrm{occ}} \mathrm{tr}\left[ P^n_{\boldsymbol{k}} \right]_{\text{no scatt}}^{\text{BZ av}}$ recovers the exact crystalline result, except close to the transition due to finite precision effects. On the other hand, the trace of the projector square $\sum_{n\in \mathrm{occ}} \mathrm{tr}\left[ \left( P^n_{\boldsymbol{k}} \right)^2 \right]_{\text{no scatt}}^{\text{BZ av}}$, which should be equal to the filling, is just slightly ($\sim 8-25\%$) smaller due to neglecting the momentum scattering.}
    \label{Sfig:qB_trP_qB_trPP}
\end{figure}

We have also verified that the correct crystalline results are obtained numerically in our bismuthene and Bi bilayer tight-binding models. Indeed, Fig.~\ref{Sfig:qB_trP_qB_trPP} shows the number of occupied states per unit cell $\sum_{n\in \mathrm{occ}} \mathrm{tr}\left[ P^n_{\boldsymbol{k}} \right]_{\text{no scatt}}^{\text{BZ av}}$ at $\boldsymbol{k}=0$ as a function of the onsite SOC for crystalline bismuthene and Bi bilayer. In both models, this number of occupied states (or filling) is constant and equal to 4 and 6, as expected, since they correspond to half-filling in bismuthene and Bi bilayer, respectively. Note that the filling artificially deviates from these values close to the topological transition. However, this is an artifact stemming from the finite KPM resolution. Indeed, this artifact only appears close to the transition, which is where the bulk gap is smaller, and therefore is where the required precision to obtain the correct results is higher. We have checked that the deviations from the exact filling shrink when increasing the KPM precision and the system size.

In summary, we have shown that, by averaging over the BZ types and multiplying by $N_{\mathrm{s/c}}$, we recover the correct values in the crystal for the quantities that involve the trace of one projector. This exact result is recovered despite neglecting both the scattering by different reciprocal lattice vectors and the non-orthogonality of the plane waves. This means that the scattering does not play a crucial role in the quantities that involve the trace of only one projector.

\subsubsection{Structural spillage without scattering in the tight-binding approximation}

Now, let us consider quantities that involve the trace of two projectors, such as the spillage. Unlike in the quantities involving just one projector, here scattering plays an important role. Indeed, we will show that scattering should be included to obtain the exact result in the crystalline limit (see, e.g., Eq.~\eqref{eq:qB-spillage-crystal}, where the sum over $\boldsymbol{G}'$ represents the scattering). However, as explained in Appendix \ref{app:spillage_TB_multi_no_scatt}, the scattering has to be neglected in order to be able to use the structural spillage to compare amorphous and crystalline systems. Nevertheless, we will also show that, even if the crystalline results are not exactly recovered, our method gives reasonably good results, which allows the structural spillage to work as a topological indicator also in the tight-binding approximation.

Consider, the trace of $\left(P^n_{\boldsymbol{k}}\right)^2$, which should be equal to one if $P^n_{\boldsymbol{k}}$ is a projector.
If we include scattering and average over Brillouin zones this exact condition is fulfilled for the crystal, as can be checked explicitly:
\begin{equation}
\begin{split}
    \mathrm{tr}\left[ \left(P^n_{\boldsymbol{k}}\right)^2 \right]_{\text{scatt}}^{\text{BZ av}} & = 
    \frac{N_{\text{s/c}}}{N_{\mathrm{BZs}}} \sum_{a \in \mathrm{BZs}} \frac{N_{\text{s/c}}}{N_{\mathrm{BZs}}} \sum_{a' \in \mathrm{BZs}} \left[ \langle \boldsymbol{k} + \boldsymbol{G}_a \big| P^n_{\boldsymbol{k}} \big| \boldsymbol{k} + \boldsymbol{G}_a + \boldsymbol{G}_{a'} \rangle \langle \boldsymbol{k} + \boldsymbol{G}_a + \boldsymbol{G}_{a'} \big| P^n_{\boldsymbol{k}} \big| \boldsymbol{k} + \boldsymbol{G}_a \rangle \right] = \\
    & = \sum_{A, B, C, D} c_{\boldsymbol{k}}^{nA} \left(c_{\boldsymbol{k}}^{nB}\right)^* c_{\boldsymbol{k}}^{nC} \left(c_{\boldsymbol{k}}^{nD}\right)^* 
    \frac{1}{N_{\mathrm{BZs}}} \sum_{a \in \mathrm{BZs}} e^{-i\boldsymbol{G}_a\cdot(\boldsymbol{t}_{AB}+\boldsymbol{t}_{CD})}
    \frac{1}{N_{\mathrm{BZs}}} \sum_{a' \in \mathrm{BZs}} e^{-i\boldsymbol{G}_{a'}\cdot\boldsymbol{t}_{CB}} = \\
    & = \sum_{A, B, D} c_{\boldsymbol{k}}^{nA} \big|c_{\boldsymbol{k}}^{nB}\big|^2 \left(c_{\boldsymbol{k}}^{nD}\right)^* 
    \frac{1}{N_{\mathrm{BZs}}} \sum_{a \in \mathrm{BZs}} e^{-i\boldsymbol{G}_a\cdot\boldsymbol{t}_{AD}} = \sum_{A, B} \big|c_{\boldsymbol{k}}^{nA}\big|^2 \big|c_{\boldsymbol{k}}^{nB}\big|^2 = 1.
\end{split}
\end{equation}
However, including scattering is not possible in general, unlike BZ averaging. As explained above, the scattering cannot be taken into account when the two projectors belong to systems with a different lattice structure. Therefore, when computing two-projector quantities we still perform the {BZ} average on the external sum over $\boldsymbol{G}_{a}$, but are forced to neglect the scattering resummation over $\boldsymbol{G}_{a'}$: 
\begin{equation}
\begin{split}
    \mathrm{tr}\left[ \left(P^n_{\boldsymbol{k}}\right)^2 \right]_{\text{no scatt}}^{\text{BZ av}} & = 
    \frac{N_{\text{s/c}}}{N_{\mathrm{BZs}}} \sum_{a \in \mathrm{BZs}} \left[ \langle \boldsymbol{k} + \boldsymbol{G}_a \big| P^n_{\boldsymbol{k}} \big| \boldsymbol{k} + \boldsymbol{G}_a \rangle \langle \boldsymbol{k} + \boldsymbol{G}_a \big| P^n_{\boldsymbol{k}} \big| \boldsymbol{k} + \boldsymbol{G}_a \rangle \right] = \\
    & = \frac{1}{N_{\text{s/c}}} \sum_{A, B, C, D} c_{\boldsymbol{k}}^{nA} \left(c_{\boldsymbol{k}}^{nB}\right)^* c_{\boldsymbol{k}}^{nC} \left(c_{\boldsymbol{k}}^{nD}\right)^* 
    \frac{1}{N_{\mathrm{BZs}}} \sum_{a \in \mathrm{BZs}} e^{-i\boldsymbol{G}_a\cdot(\boldsymbol{t}_{AB}+\boldsymbol{t}_{CD})} = \\
    & = \frac{1}{N_{\text{s/c}}} \sum_{A, B, C, D} c_{\boldsymbol{k}}^{nA} \left(c_{\boldsymbol{k}}^{nB}\right)^* c_{\boldsymbol{k}}^{nC} \left(c_{\boldsymbol{k}}^{nD}\right)^* 
    \delta_{\boldsymbol{t}_{AB}+\boldsymbol{t}_{CD},0}.
\end{split}
\label{eq:trPP_avBZ}
\end{equation}
Although this equation does not exactly recover the crystalline value, we have numerically verified that the sum over occupied bands of this Eq.~\eqref{eq:trPP_avBZ}, $\sum_{n\in\mathrm{occ}} \mathrm{tr}[ (P^n_{\boldsymbol{k}})^2]_{\text{no scatt}}^{\text{BZ av}}$, gives values just $\sim 8 - 25 \%$ smaller than $\sum_{n\in\mathrm{occ}} \mathrm{tr}[ P^n_{\boldsymbol{k}}]_{\text{no scatt}}^{\text{BZ av}}$ in the crystal, as shown in Fig.~\ref{Sfig:qB_trP_qB_trPP}. Therefore, we take this as a reasonable approximation, especially taking into account that this quantity can also be computed when one of the projectors corresponds to an amorphous structure. Applying this method to the structural quasi-Bloch spillage, we arrive at Eq.~\eqref{eq:qB-spillage-no-scatt-matrix-TB}.

In order to implement the tight-binding spillage of Eq.~\eqref{eq:qB-spillage-no-scatt-matrix-TB} we need to account for a final detail: the choice of a representative BZ of each type. This is a requirement because we introduced the average over BZ types in Eqs.~\eqref{eq:trP_avBZ}-\eqref{eq:trPP_avBZ}. To perform this average, one has to select one representative for each type of BZ. To this end, let us consider the example of the honeycomb lattice relevant to our Bi models, which has $N_{\mathrm{BZs}}=3$ types of BZ, as sketched in Fig.~\ref{Sfig:BZ_types_honeycomb}. Due to the argument which lead us to Eq.~\eqref{eq:qB-spillage-no-scatt-matrix-TB-single-site} in Appendix \ref{app:spillage_TB_single_site_subsec}, the optimal criterium for choosing the BZ representatives is to consider the ones whose reciprocal lattice vector is smaller in modulus. For example, the first BZ will always be chosen as the representative of the BZs characterized by a phase $e^{i\boldsymbol{G}\cdot\boldsymbol{t_{AB}}} = 1$. There can still be several options, such as the three possibilities for the BZs with phases $e^{i\boldsymbol{G}\cdot\boldsymbol{t_{AB}}} = e^{\pm i 2\pi/3}$. In this case, one can choose any of them. A better choice however is to perform an angular average over them. Indeed, while the crystal is anisotropic, the amorphous structure is effectively isotropic. In particular, although the total traces in the crystal are exactly the same in all equivalent BZs, some orbital-resolved quantities might vary. For instance, in the honeycomb lattice, if the occupied eigenstate at $\boldsymbol{G} = 4\pi/\sqrt{3} (0,1)$ is of $p_y$ character, the eigenstate at the threefold rotated $\hat{C}_3 \boldsymbol{G} = 4\pi/\sqrt{3} (-\sqrt{3}/2,-1/2)$ is of the threefold rotated $-(\sqrt{3}/2) p_x - (1/2) p_y$ character. On the other hand, for sufficiently large samples, amorphous structures are expected to be isotropic in momentum space. Therefore, one would ideally perform an angular average over the $\boldsymbol{G}$ corresponding to equivalent BZs with the same modulus, but pointing in a different direction. In the honeycomb lattice, the quantity corresponding to the BZs with phase $e^{i\boldsymbol{G}\cdot\boldsymbol{t_{AB}}} = e^{+ i 2\pi/3}$ would be an average over the three BZs shown in blue in Fig.~\ref{Sfig:BZ_types_honeycomb}. Consequently, when the corresponding crystal displays a honeycomb lattice, the angle-averaged Eq.~\eqref{eq:qB-spillage-no-scatt-matrix-TB} for the structural quasi-Bloch spillage in the tight-binding approximation reads:
\begin{equation}
    \gamma_{\mathrm{qB}}^{\mathrm{TB}}(\boldsymbol{k}) = \frac{2}{3} \left\{ 
    \frac{1}{2} \text{tr} \left[ \left( P_{\boldsymbol{k}+\boldsymbol{G}_0} - \tilde{P}_{\boldsymbol{k}+\boldsymbol{G}_0} \right)^2 \right] + 
    \frac{1}{3} \sum_{\boldsymbol{G}_1^m} \frac{1}{2} \text{tr} \left[ \left( P_{\boldsymbol{k}+\boldsymbol{G}_1^m} - \tilde{P}_{\boldsymbol{k}+\boldsymbol{G}_1^m} \right)^2 \right] + 
    \frac{1}{3} \sum_{\boldsymbol{G}_2^m} \frac{1}{2} \text{tr} \left[ \left( P_{\boldsymbol{k}+\boldsymbol{G}_2^m} - \tilde{P}_{\boldsymbol{k}+\boldsymbol{G}_2^m} \right)^2 \right] \right\},
\label{eq:qB_spillage_TB_av_angles}
\end{equation}
where:
\begin{alignat}{2}
    & \boldsymbol{G}_0 = 0 &&\Rightarrow e^{-i\boldsymbol{G}_0 \cdot \boldsymbol{t}_{AB}} = 1, \\
    & \begin{Bmatrix}
        \boldsymbol{G}_1^0 = 4\pi/\sqrt{3} (0,1) \\
        \boldsymbol{G}_1^1 = \hat{C}_3 \boldsymbol{G}_1^0 = 4\pi/\sqrt{3} (-\sqrt{3}/2,-1/2)
        \\\boldsymbol{G}_1^2 = (\hat{C}_3)^2 \boldsymbol{G}_1^0 = 4\pi/\sqrt{3} (\sqrt{3}/2,-1/2)
    \end{Bmatrix} &&\Rightarrow e^{-i\boldsymbol{G}_1^m \cdot \boldsymbol{t}_{AB}} = e^{i2\pi/3}, \\
    & \begin{Bmatrix}
        \boldsymbol{G}_2^0 = 4\pi/\sqrt{3} (0,-1) \\
        \boldsymbol{G}_2^1 = \hat{C}_3 \boldsymbol{G}_2^0 = 4\pi/\sqrt{3} (\sqrt{3}/2,1/2)
        \\\boldsymbol{G}_2^2 = (\hat{C}_3)^2 \boldsymbol{G}_2^0 = 4\pi/\sqrt{3} (-\sqrt{3}/2,1/2)
    \end{Bmatrix} &&\Rightarrow e^{-i\boldsymbol{G}_2^m \cdot \boldsymbol{t}_{AB}} = e^{-i2\pi/3}.
\end{alignat}
Eq.~\eqref{eq:qB_spillage_TB_av_angles} is a specific instance of the general Eq.~\eqref{eq:qB-spillage-no-scatt-matrix-TB} that we used for computing the spillage in our bismuthene and Bi bilayer tight-binding models. However, we have also checked that in these models, for the system sizes considered, performing the angular average or not does not noticeably change the results. 

In summary, our proposed method for computing two-projector quantities, such as the structural spillage, consists of neglecting the momentum scattering, performing an average over the different types of BZs, and multiplying by the number of sites per unit cell in the corresponding crystal. Applying this method to the structural quasi-Bloch spillage, we arrive at the final expression for the structural spillage in the tight-binding approximation, Eq.~\eqref{eq:qB-spillage-no-scatt-matrix-TB} of the main text. To conclude, we highlight that, in the specific case when the number of types of BZs is infinite or very large,  \eqref{eq:qB-spillage-no-scatt-matrix-TB} would involve reciprocal lattice vectors $|\boldsymbol{G}| \gg 2\pi/a$, with $a$ the crystalline lattice constant. In this case, as in the single-site case, we may introduce a momentum cutoff and consider only the reciprocal lattice vectors $\boldsymbol{G}$ smaller than this cutoff.

\subsection{Phase transition criterion in the tight-binding approximation}\label{app:spillage_TB_criterion}

In this section we define our criterion to choose the topological transition. To this end it is important to note first that, as mentioned above, Eq.~\eqref{eq:qB-spillage-no-scatt-matrix-TB} does not exactly recover the values of the Bloch spillage when applied to two crystals with and without SOC, because we neglected scattering. However, we have numerically verified that it results in similar values. In particular, the maximum spillage without scattering is $\mathrm{max}\left[\gamma_{\mathrm{qB}}^{\mathrm{TB}}(\boldsymbol{k}=0)\right] = 1.5$ in the two models, which is a factor of $4/3$ smaller than the exact spillage $\mathrm{max}\left[\gamma_{\mathrm{qB}}(\boldsymbol{k}=0)\right] = 2$ that would be recovered after considering the scattering. This is related to the fact that $\sum_{n\in \mathrm{occ}} \mathrm{tr}\left[ \left( P^n_{\boldsymbol{k}=0} \right)^2 \right]_{\text{no scatt}}^{\text{BZ av}}$ is a factor of $4/3$ smaller than $\sum_{n\in \mathrm{occ}} \mathrm{tr}\left[ P^n_{\boldsymbol{k}=0} \right]_{\text{no scatt}}^{\text{BZ av}}$ in the topological and trivial phases for the bismuthene and Bi bilayer tight-binding models, respectively (see Fig.~\ref{Sfig:qB_trP_qB_trPP}). There is no reason to believe that this factor is universal, and thus we consider it model dependent.

With this in mind, in order to identify the topological phases in a tight-binding phase diagram, we take the criterion that the topological transition occurs when the quasi-Bloch spillage of Eq.~\eqref{eq:qB-spillage-no-scatt-matrix-TB} equals to half the maximum value of the spillage between two topologically different crystals when scattering is neglected. In both our models, this critical value equals 0.75. However, in general, this critical value of the tight-binding structural spillage will be model-dependent, and must be determined in a case-to-case basis.

\section{\label{app:pw-spillage} Absence of a corresponding crystal: spin-orbit plane-wave spillage}

One of our assumptions for applying the structural quasi-Bloch spillage of Eqs.~\eqref{eq:qB-spillage}-\eqref{eq:qB-spillage-no-scatt-matrix-TB} is that there exists a crystalline structure with similar local environments to the non-crystalline one. While this is a quite generic feature \cite{Zallen}, there are also some amorphous and quasicrystalline structures whose local environment is different to any crystalline phase of the same material. In this case, while the structural quasi-Bloch spillage could still be calculated, it would probably not be very indicative of the topology, since many possibly trivial band inversions could occur. 

In this case, one could again resort to computing the spin-orbit Bloch spillage comparing an amorphous supercell with and without SOC, as proposed for crystals by Liu and Vanderbilt \cite{liu_spin-orbit_2014}. However, as mentioned in the main text, this would always be a large quantity due to the big size of the supercell. Liu and Vanderbilt proposed to fix this issue by analyzing valence- and conduction-band-resolved spillages. However, these are not gauge-invariant, and a careful analysis is required to discern the topological character using this method. These solutions are not practical from the point of view of a performing high-throughput screening of amorphous materials, where it is desirable to define a quantity that is easily implemented and analyzed using ab-initio codes. 

For such cases without a crystalline counterpart, we propose instead a plane-wave-resolved spin-orbit spillage comparing an amorphous system with and without SOC. This spin-orbit plane-wave spillage $\gamma_{\mathrm{pw}}(\boldsymbol{p})$ is defined as in Eq.~\eqref{eq:qB-spillage-general} but without the sum over crystalline reciprocal lattice vectors $\boldsymbol{G}$:
\begin{align}
	\gamma_{\mathrm{pw}}(\boldsymbol{p}) & = \frac{1}{2} \sum_{\boldsymbol{p}'} \sum_{\alpha\beta} \left[ P^{\alpha\beta}_{\boldsymbol{p},\boldsymbol{p}'} P^{\beta\alpha}_{\boldsymbol{p}',\boldsymbol{p}} - P^{\alpha\beta}_{\boldsymbol{p},\boldsymbol{p}'} \tilde{P}^{\beta\alpha}_{\boldsymbol{p}',\boldsymbol{p}} \right] + \left[P \leftrightarrow \tilde{P} \right]
	\label{eq:pw-spillage}
\end{align}
where $\boldsymbol{p}$ and $\boldsymbol{p}'$ are plane-wave momenta. For a supercell Gamma calculation in DFT,  $\boldsymbol{p}$ and $\boldsymbol{p}'$ would be the supercell reciprocal lattice vectors. Now, since both systems that are being compared have the same structure, Eq.~\eqref{eq:pw-spillage} can also be applied within a tight-binding approximation. However, for the latter approximation, one could first compute the much more efficient plane-wave spillage without scattering, which would read:
\begin{equation}
    \gamma_{\mathrm{pw}}^{\text{no scatt}}(\boldsymbol{p}) = \frac{1}{2} \text{tr} \left[ \left( P_{\boldsymbol{p}} - \tilde{P}_{\boldsymbol{p}} \right)^2 \right],
\label{eq:pw-spillage-no-scatt-matrix-TB}
\end{equation}
We however leave the benchmarking of the plane-wave spillage for future work.

\end{document}